\documentclass[a4paper,fleqn,usenatbib]{mnras}

\usepackage{mathptmx}
\usepackage{amsmath}
\usepackage{amssymb}
\usepackage{txfonts}

\usepackage[T1]{fontenc}
\usepackage{ae,aecompl}

\usepackage{graphicx}	
\usepackage{amsmath}	
\usepackage{amssymb}	

\newcommand{\photu}{ph\ cm$^{-2}$ s$^{-1}$ sr$^{-1}$ \AA$^{-1}$}
\newcommand{\galex}{{\it GALEX}}

\title[Diffuse Galactic Light]{Modelling Dust Scattering in our Galaxy}

\author[Jayant Murthy]{
Jayant Murthy,$^{1}$\thanks{E-mail: jmurthy@yahoo.com}
\\
$^{1}$Indian Institute of Astrophysics, Bangalore 560 034\\
}

\date{Accepted XXX. Received YYY; in original form ZZZ}

\pubyear{2015}

\begin{document}
\label{firstpage}
\pagerange{\pageref{firstpage}--\pageref{lastpage}}
\maketitle

\begin{abstract}

I have used Monte Carlo models with multiple scattering to predict the dust scattered light from our Galaxy and have compared the predictions with data in two UV bands from the \galex\ spacecraft. I find that 90\% of the scattered light arises from less than 1000 stars with 25\% from the 10 brightest. About half of the diffuse radiation originates within 200 pc of the Sun with a maximum distance of 600 pc. Multiple scattering is important at any optical depth with 30\% of the flux being multiply scattered even at zero reddening. I find that the global distribution of the scattered light is insensitive to the dust distribution with grains of $0.3 < a < 0.5$ and $g < 0.6$. There is an offset between the model and the data of 100 and 200 \photu\ in the FUV and NUV, respectively, at the poles rising to 200 --- 400 \photu\ at lower latitudes. 

The Monte Carlo code and the models of diffuse radiation for different values of the optical constants are available for download. 

\end{abstract}




\section{Introduction}

The diffuse ultraviolet (UV) background was first detected by \citet{Hayakawa1969} but its faintness and the need for space-borne observations rendered progress slow for the next few decades (reviewed by \citet{Bowyer1991} and \citet{Henry1991} with a recent review by \citet{Murthy2009}). There have been three large scale surveys of the diffuse ultraviolet sky over the last two decades. The first was the NUVIEWS (Narrowband Ultraviolet Imaging Experiment for Wide-Field Surveys) rocket flight \citep{Schiminovich2001} which observed about 25\% of the sky and noted the strong enhancement toward the Galactic plane with other bright spots near star forming regions such as Ophiuchus. More recently, the SPEAR/FIMS mission  carried out a spectroscopic UV survey of the sky \citep{Edelstein2006} and the Galaxy Evolution Explorer (\galex) observed about 75\% of the sky in two ultraviolet bands \citep{Murthy2014}.

The greatest part of the diffuse background, particularly at low latitudes, is thought to be due to starlight scattered by interstellar dust grains \citep{Draine2003} and models of the background have taken the incoming starlight and scattered it from the grains with different assumptions for the distribution of the exciting stars and the scattering dust \citep{Jura1979, Murthy1995, Gordon2001, Schiminovich2001, Seon2015}. The availability of the all-sky \galex\ data provides a powerful incentive to revisit the nature of the diffuse background with a consistent set of data and a unified modelling approach. Such deviations from dust scattered predictions may have important consequences for our understanding of the evolution of galaxies and the dynamical history of the Universe \citep{Overduin2004}. I will describe here a new model of the dust scattered background and its application to the \galex\ observations in its two ultraviolet bands. 

\section{Data}

The \galex\ spacecraft and its mission has been described by \citet{Martin2005} and \citet{Morrissey2007}. The spacecraft was operational from 2003 June 7 until 2013 June 28 and observed most of the sky with a resolution of 5 -- 10\arcsec\ in two UV bands. The FUV (1521 \AA) detector was plagued with intermittent failures of the high voltage power supply (HVPS) and finally ceased to work after 2009 May while the NUV (2361 \AA) instrument took data until the spacecraft was shut down. Most of the observations were made at high Galactic latitudes to avoid damage to the detectors from the bright diffuse background but there was a concerted effort to map brighter regions, including at low Galactic latitudes, near the end of the mission. Unfortunately, the  FUV detector had already failed by this time so there are very few FUV observations at low latitudes.

\begin{figure}
\includegraphics[width=3.5in]{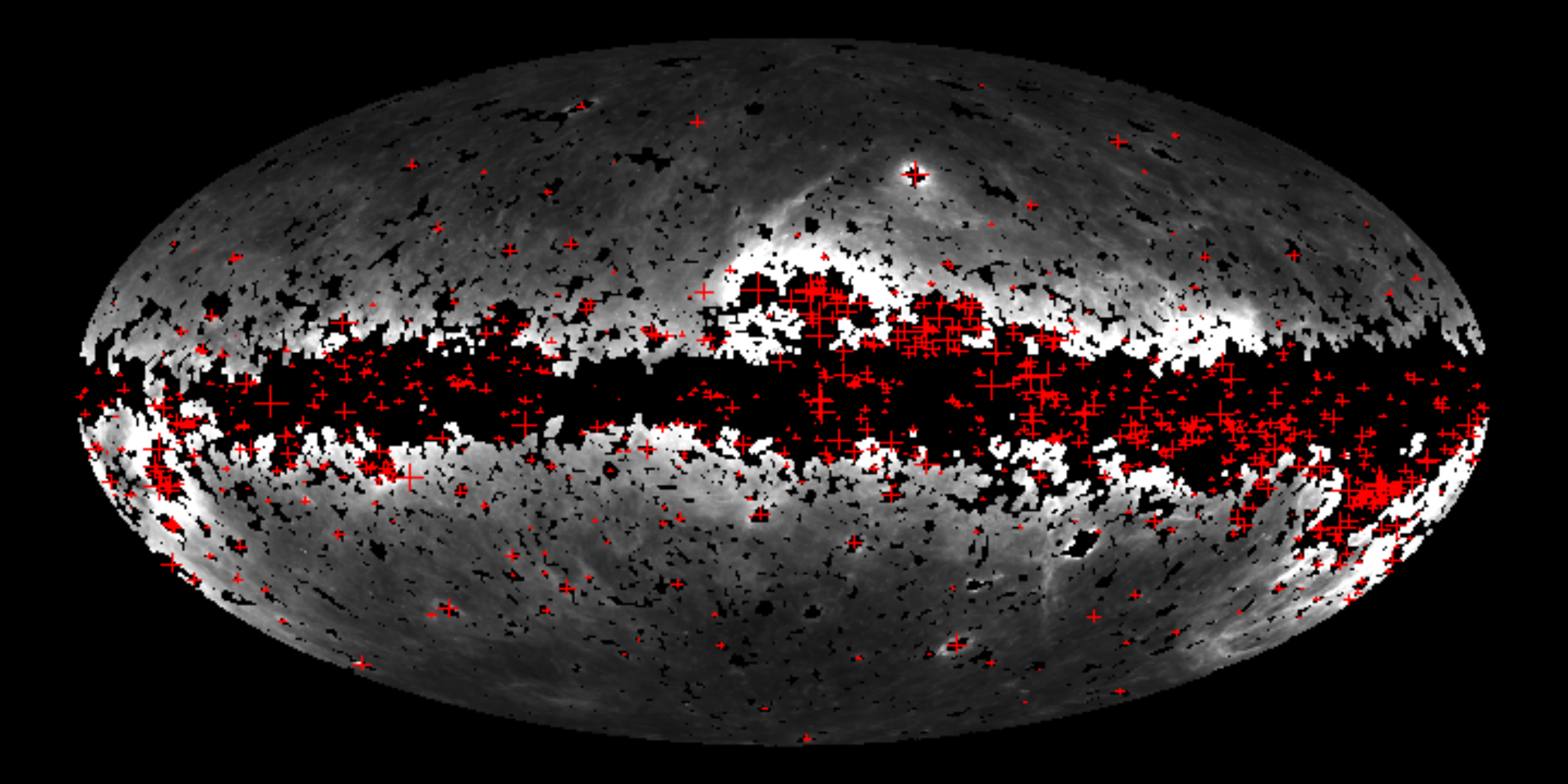}
\includegraphics[width=3.5in]{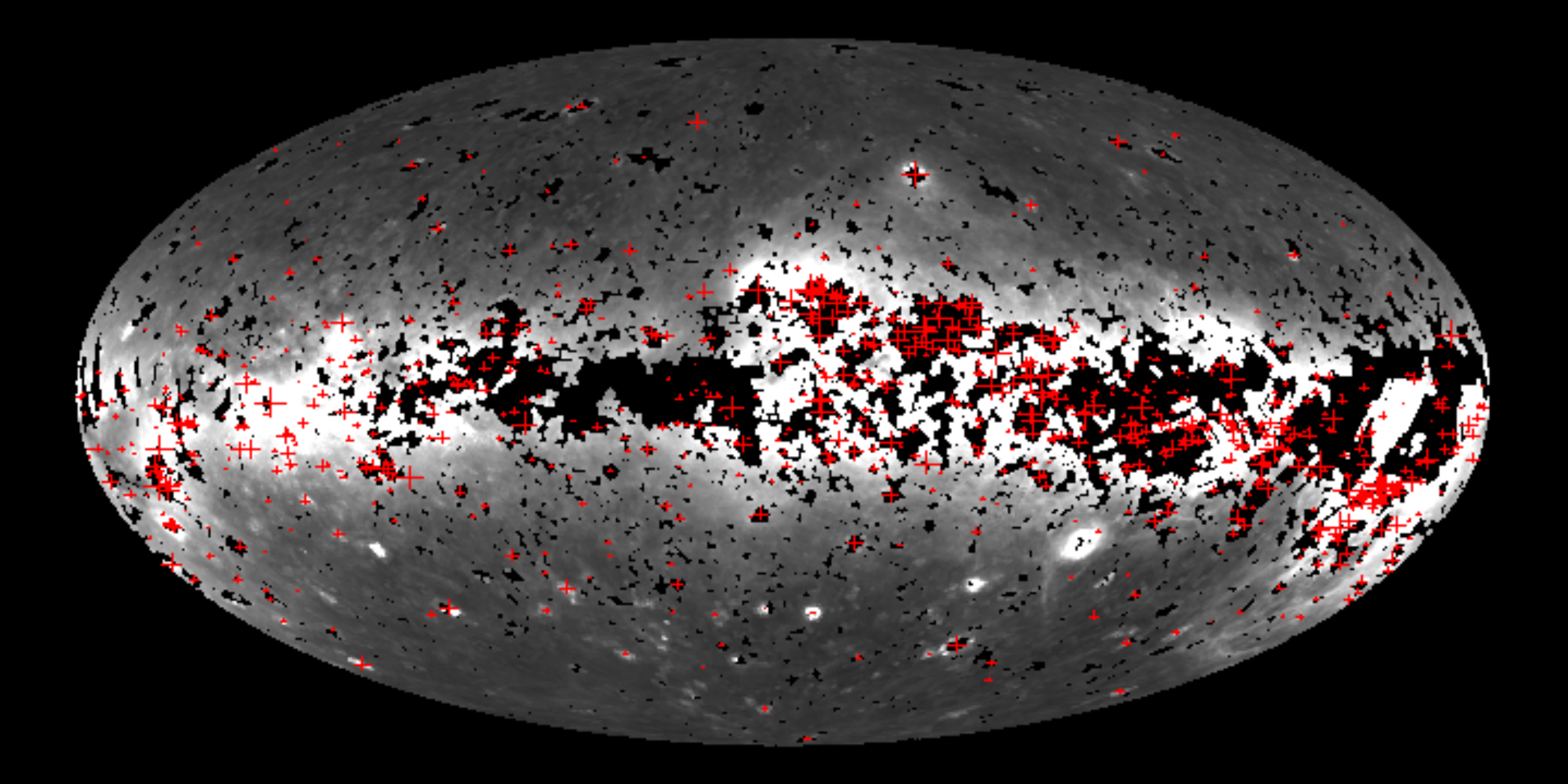}
\caption{Observed fluxes in the FUV (top) and NUV (bottom) \galex\ bands. The brightest stars are plotted as + symbols with the width of the symbol proportional to the log of the brightness at the respective wavelength. Black areas were not observed by \galex.}
\label{obs_stars}
\end{figure}

\citet{Murthy2014} used the \galex\ data to construct maps of the diffuse background in both the  FUV and NUV bands at $0.1^{\circ}$ spatial resolution with the foreground emission (airglow and zodiacal light) subtracted (Fig. \ref{obs_stars}).  The 1000 brightest stars are overplotted on the images as + signs with the size of the symbol proportional to the log of the brightness. Gould's Belt is prominent in both bands as are halos around a number of bright stars \citep{Murthy2011}. A full description of the methodology in the production of these maps is given in the paper and the maps are available from the High Level Science Products (HLSP) data repository\footnote{https://archive.stsci.edu/prepds/uv-bkgd/} at the Space Telescope Science Institute.

\section{Modelling}

The radiative transfer problem in galaxies has been reviewed by  \citet{Steinacker2013} with the much simpler problem of scattering only addressed in their Section 5.1. The problem can be broken into three parts: the stellar distribution; the dust distribution; and the scattering function. Photons are emitted by the stars and are scattered by the interstellar dust to the detector. The scattering function is commonly assumed to be the Henyey-Greenstein function \citep{Henyey1941} which is dependent on two free parameters: the albedo or reflectivity ($a$) and the phase function asymmetry factor ($g \equiv < cos(\theta) >$), where $g = 0$ implies that the grains scatter isotropically and $g = 1$ implies fully forward scattering grains. \citet{Draine2003} has pointed out that the scattering may be more complex with a possible reverse scattering component but the data have not yet been good enough to support added complexity in the scattering function.

\citet{Henry1977} showed that the interstellar radiation field in the UV could be calculated through an integration over the stars in a standard catalog. In this work, I have used the Hipparcos catalog \citep{Perryman1997} which contains over 100,000 stars with the spectral type, B and V magnitude and distance of each star. I modelled the spectrum of each star using template spectra from \citet{Castelli2004} with the translation from spectral type to model number as per their instructions\footnote{http://www.stsci.edu/hst/observatory/crds/castelli\_kurucz\_atlas.html}. I then calculated the observed E(B - V) from the cataloged B and V magnitudes and the intrinsic (B - V) and finally the unreddened flux from each star assuming the Milky Way extinction curve of \citet{Draine2003}. This is the source function in my model: the number of photons from each star at the wavelength of interest.

\begin{figure}
\includegraphics[width=4in]{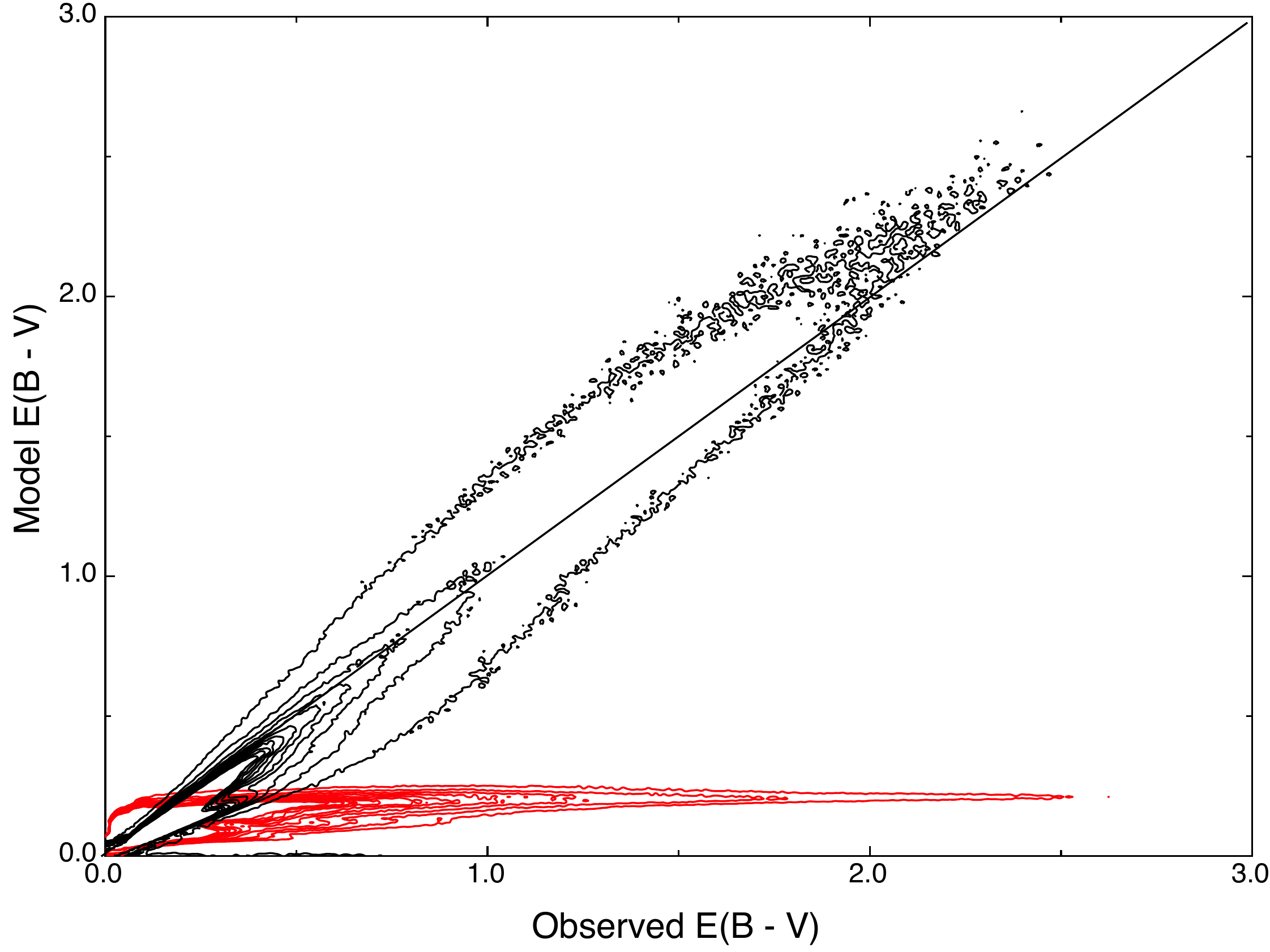}
\caption{The correlation between the E(B-V) from \citet{Schlegel1998} and the modelled E(B-V) is shown as a density plot for Model 1 (red lines) and Model 2.}
\label{pred_model_ebv}
\end{figure}

I have tried two dust distributions in order to explore their effects on the scattered light. In the first (Model 1), I have assumed that the gas density at the Galactic plane (n[H]) is 1 atom cm$^{-3}$, independent of longitude. In the second (Model 2), I scaled the dust to match the observed E(B - V) from \citet{Schlegel1998} in Model 2. In both cases, I assumed that there was a cavity of radius 80 pc around the Sun, corresponding to the Local Bubble \citep{Welsh2010}, and that the dust fell off from the Galactic plane with a scale height of 125 pc \citep{Marshall2006}. I have plotted the correlation between the observed and the modelled E(B - V) for the two models in Fig. \ref{pred_model_ebv}. My purpose in implementing these different models was to explore the factors affecting dust scattering rather than to accurately represent the distribution of interstellar dust.

A full radiative transfer model is complex because it is non-linear, non-local and multiwavelength in its formulation \citep[reviewed by][]{Steinacker2013}. However, I am only concerned with the scattering of photons at single wavelengths which is a much simpler problem \citep[Section 5.1]{Steinacker2013}. I have written a set of routines in ANSI C which are available from the ASCL \citep{Murthy2015}. The program is intended for use in the UV where the source distribution is well characterized and the volume of space is limited because of the high optical depth to UV radiation. However, the code is modular and documented and may be freely modified for other purposes.

My code is very similar to other Monte Carlo programs to predict the scattered light (eg. \citet{Wood1999, Gordon2001}). I will illustrate the program flow in my implementation by following a single photon through its multiple scatterings. I generate a new random number from a uniform distribution in each step below except in Steps 2 and 5 where two numbers are needed to determine the direction of the photon. I have used the genunf (generate a random number from a uniform distribution) function from the randlib library\footnote{http://hpux.connect.org.uk/hppd/hpux/Maths/Misc/randlib-1.3/readme.html} to generate the random numbers and, if necessary, weighted the distribution to match the desired probabilities.

\begin{enumerate}
\item A photon is emitted from a random star, where the probability of selecting a given star is weighted by the relative number of photons from that star, in a random direction. Each photon begins with a unit weight which will be reduced at each scattering.
\item Two random numbers are generated to calculate the direction of motion, one for theta (in the range from 0 --- $\pi$, measured from the z axis) and one for phi (in the range from 0 ---  $2\pi$). The position of the star is known (from the Hipparcos catalog) in Cartesian coordinates (x, y, z) and the angles are converted into a direction vector assuming a step size of 1 bin.
\item Another random number is generated to determine the optical depth the photon travels. I have divided the Galaxy into 1000 x 1000 x 1000 cells with a side of 1 pc and filled each cell with dust as per the individual model. The cross-section of the dust was taken from \citet{Draine2003}, which is a parametrization of the canonical extinction curve with R $(= A_{V} / E(B - V)) = 3.1$, the so-called Milky Way dust. I then follow the photon along until the cumulative optical depth along the path exceeds the predetermined value. This yields the Cartesian coordinates x, y, and z of the scattering location.
\item I use the ``effective peeling'' technique \citep{Zadeh1984} to send a fraction of the photon back to the detector and subtract this (small) amount of energy from the effective weight of the photon. The detector in this case is assumed to observe the entire sky with an angular resolution of 0.1$^{\circ}$ per square bin.
\item The effective weight of the photon is reduced by the albedo and a new scattering direction is determined as in step 2. The z direction is now the original direction of motion with the change of reference back to the original Cartesian coordinates using a rotation angle matrix.
\item The procedure is repeated from step 3 until the effective weight of the photon drops below a predetermined factor or the number of scatterings exceeds a specified limit ({\it nscatter}). Note that single scattering corresponds to $nscatter = 0$. In that case, the photon stops at the first interaction but the effective peeling method results in a flux at the detector.
\end{enumerate}

\begin{figure}
\includegraphics[width=3.3in]{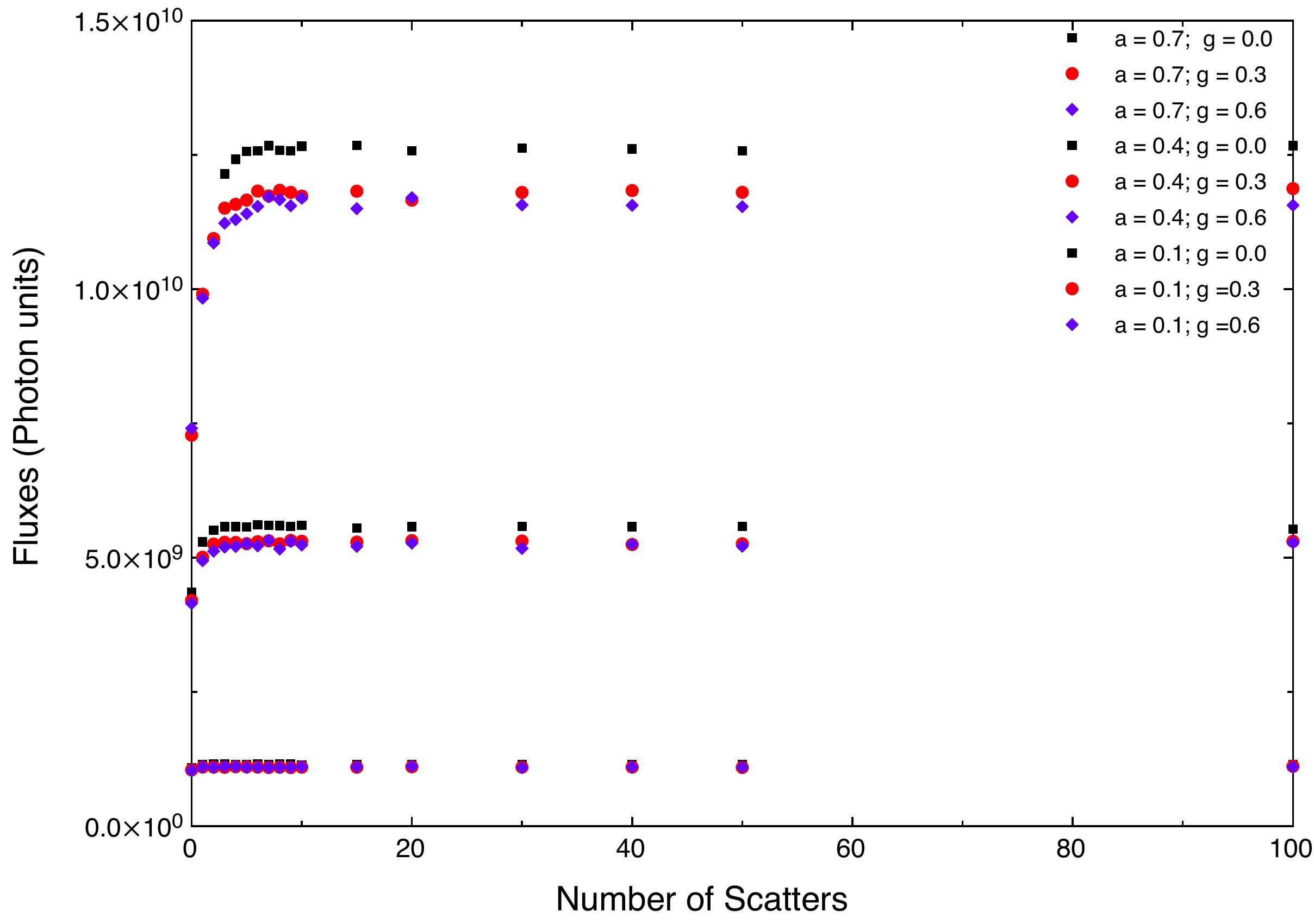}
\includegraphics[width=3.3in]{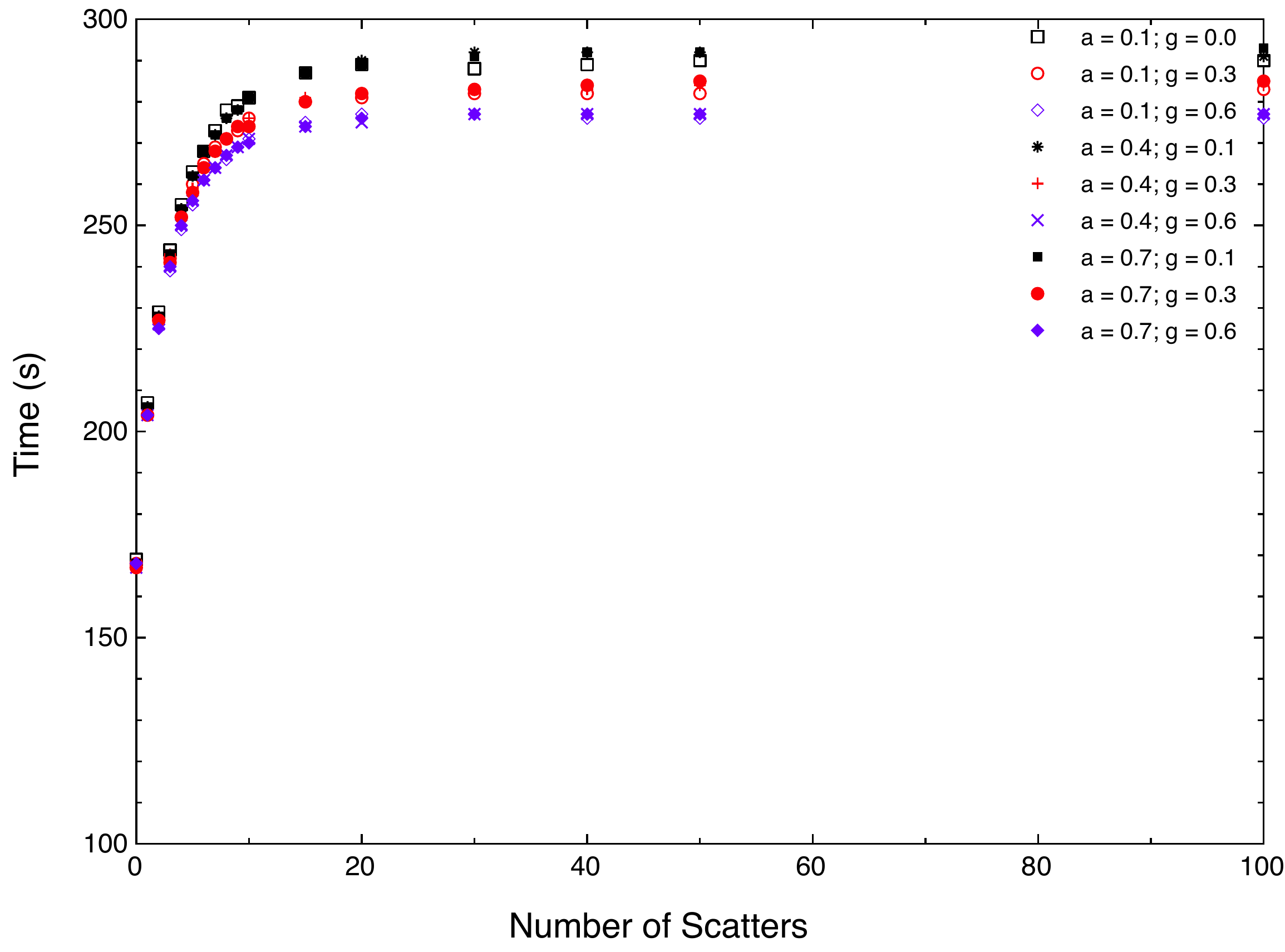}
\caption{The total flux (\photu) in the simulation is shown on the top and the amount of time per 10 million input photons (on a 3.7 GHz Quad-Core Intel Xeon E5 MacPro) is shown on the bottom. In both, the maximum number of scatterings per input photon before terminating the run is plotted on the x axis. Each point represents a run of 10 million input photons with different value of $a$ and $g$.}
\label{flux_per_10_mil}
\end{figure}

\section{Results}

I have plotted the total flux in the Galaxy at 1500 \AA\ as a function of the number of scatterings for different values of the optical constants along with the time taken for each run in Fig. \ref{flux_per_10_mil}. The flux saturates at about 5 scatterings per photon and I have therefore capped the number of scatters at that level leading to a significant savings in execution time without affecting the total flux. These numbers are from runs at 1500 \AA\ with the dust distributed as in Model 2; similar results are obtained at 2300 \AA\ and for Model 1.

\begin{figure}
\centering
\includegraphics[width=3.5in]{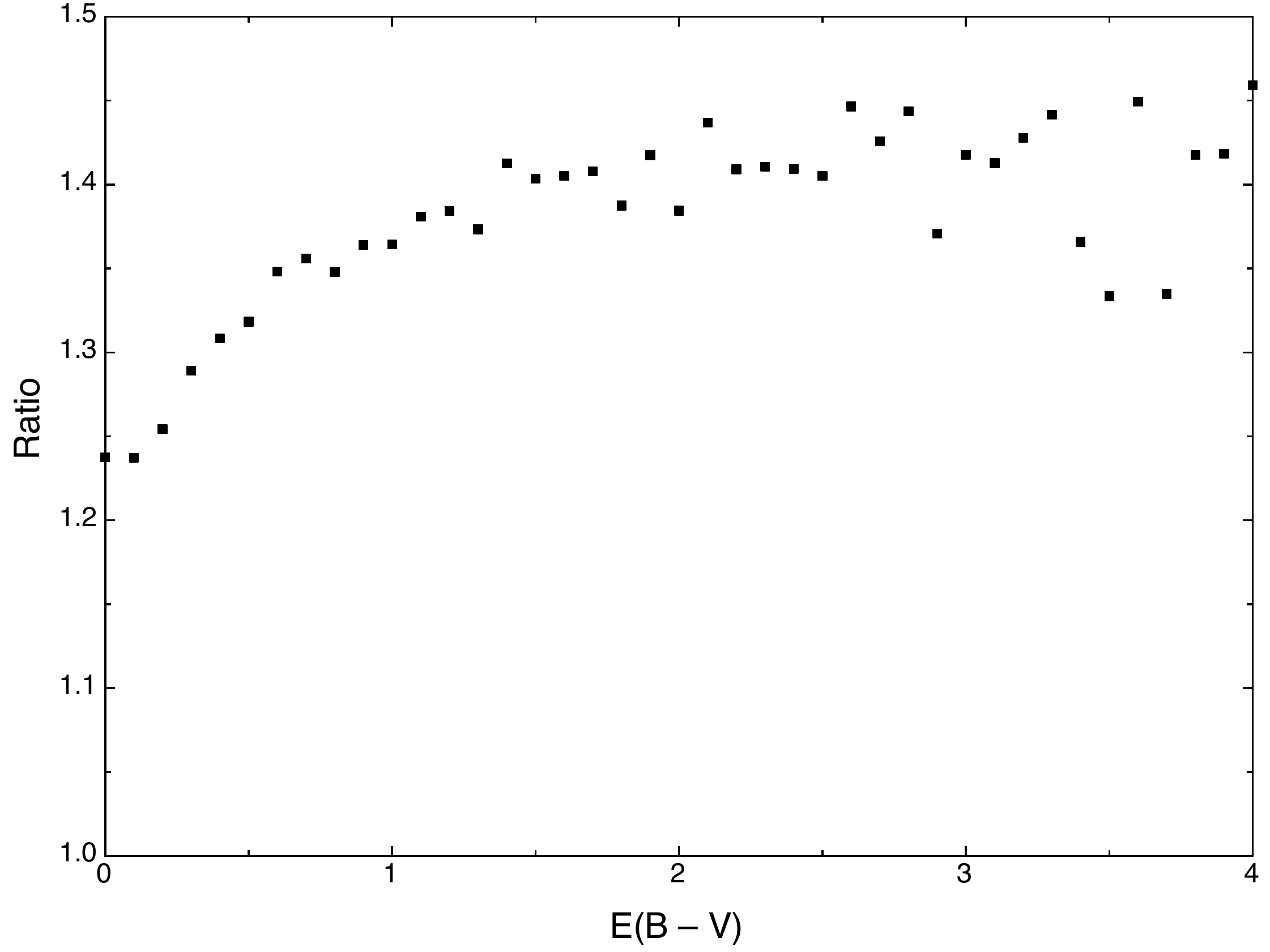}
\caption{Ratio between multiple and single scattering as a function of E(B - V) with $a = 0.4$ and $g = 0.3$.}
\label{single_multiple_ratio}
\end{figure}

It is tempting to assume that single scattering provides a reasonable estimate of the diffuse flux, especially in regions of low optical depth \citep{Murthy1995, Henry2015} because the solution may be derived exactly without recourse to Monte Carlo methods. I have plotted the ratio between the two for $a = 0.4$ and $g = 0.3$ in Fig. \ref{single_multiple_ratio}. Even at the lowest reddening,  multiple scattering gives about 25\% more flux rising to about 40\% more by E(B - V) = 1.5, although the exact value will depend on the local geometry between the stars and the dust. Much of this excess is simply because the photon still carries energy after the first interaction which is disregarded in the single scattering assumption.

\begin{figure}
\centering
\begin{minipage}[t]{0.4\textwidth}
\includegraphics[width=3.5in]{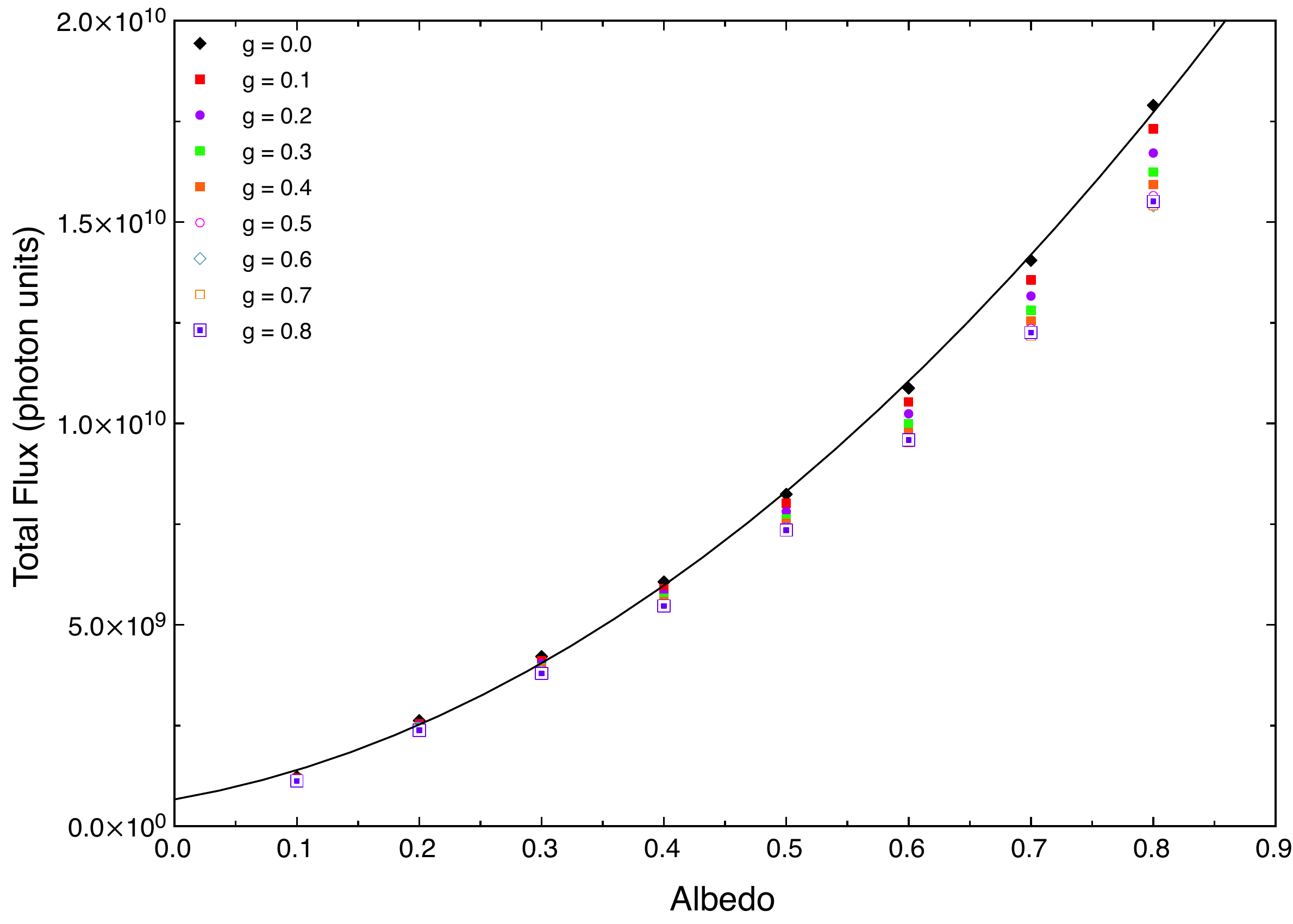}
\caption{The total flux over the entire sky is shown as a function of albedo for a range of $g$. The line represents a quadratic fit to the $g = 0$ points and is shown for comparison only.}
\label{flux_with_albedo}
\end{minipage}
\hfill
\begin{minipage}[t]{0.4\textwidth}
\includegraphics[width=3.3in]{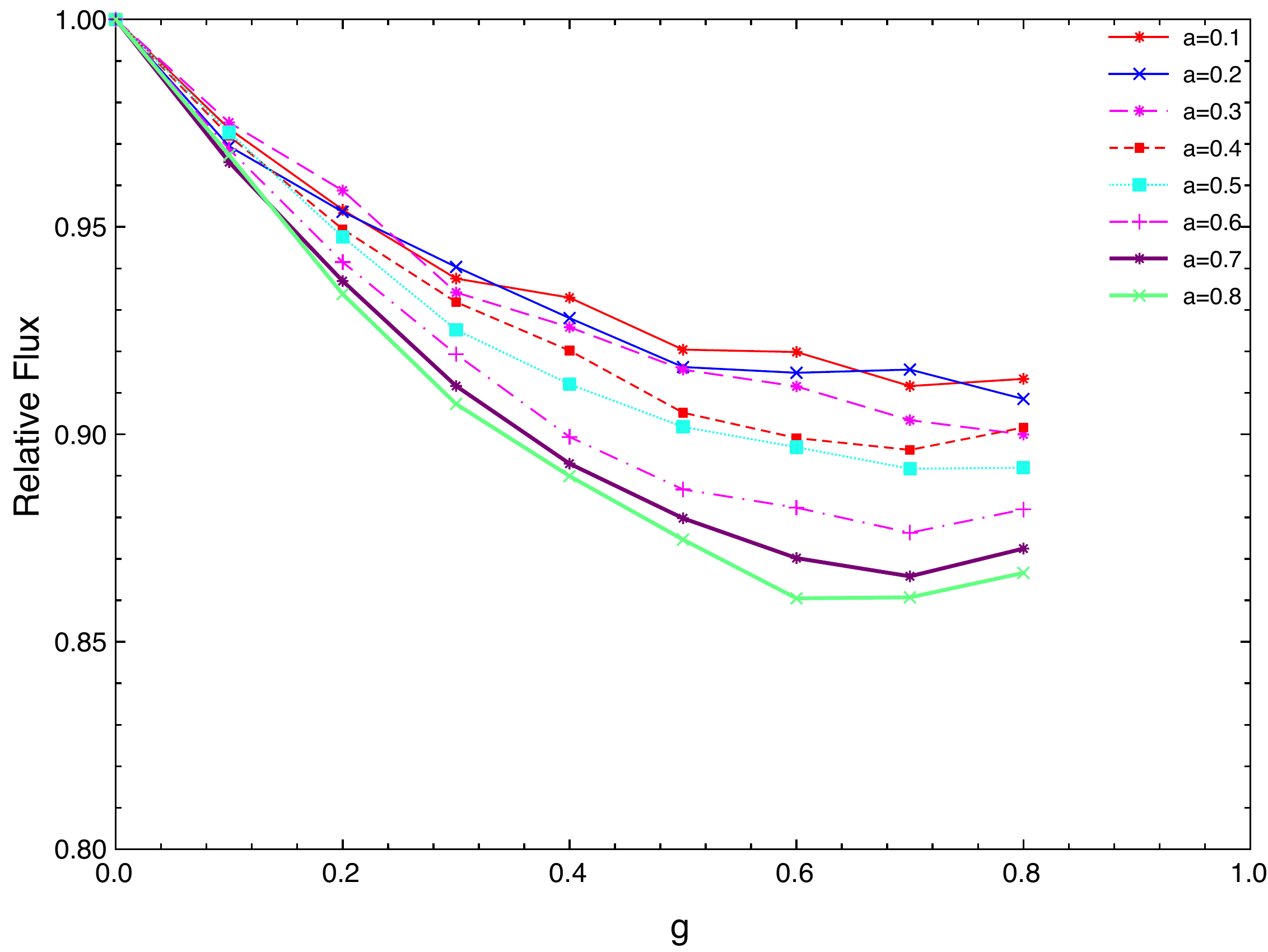}
\caption{The flux over the entire sky is shown as a function of $g$ for different values of the albedo (see legend). The fluxes for each albedo have been divided by the flux at $g = 0$.}
\label{flux_with_g}
\end{minipage}
\end{figure}

I have added the flux over the entire model sky for each combination of the optical constants in Fig. \ref{flux_with_albedo} and \ref{flux_with_g}. Unlike the single scattering case where the total flux should rise linearly with the albedo, the total flux rises as approximately the square of the albedo when multiple scattering is taken into account. The flux decreases with increasing $g$ as the photon is more likely to stay within the Galaxy for isotropic scattering and there are more scatterings per photon.

\begin{figure}
\centering
\begin{minipage}[t]{0.4\textwidth}
\includegraphics[width=3in]{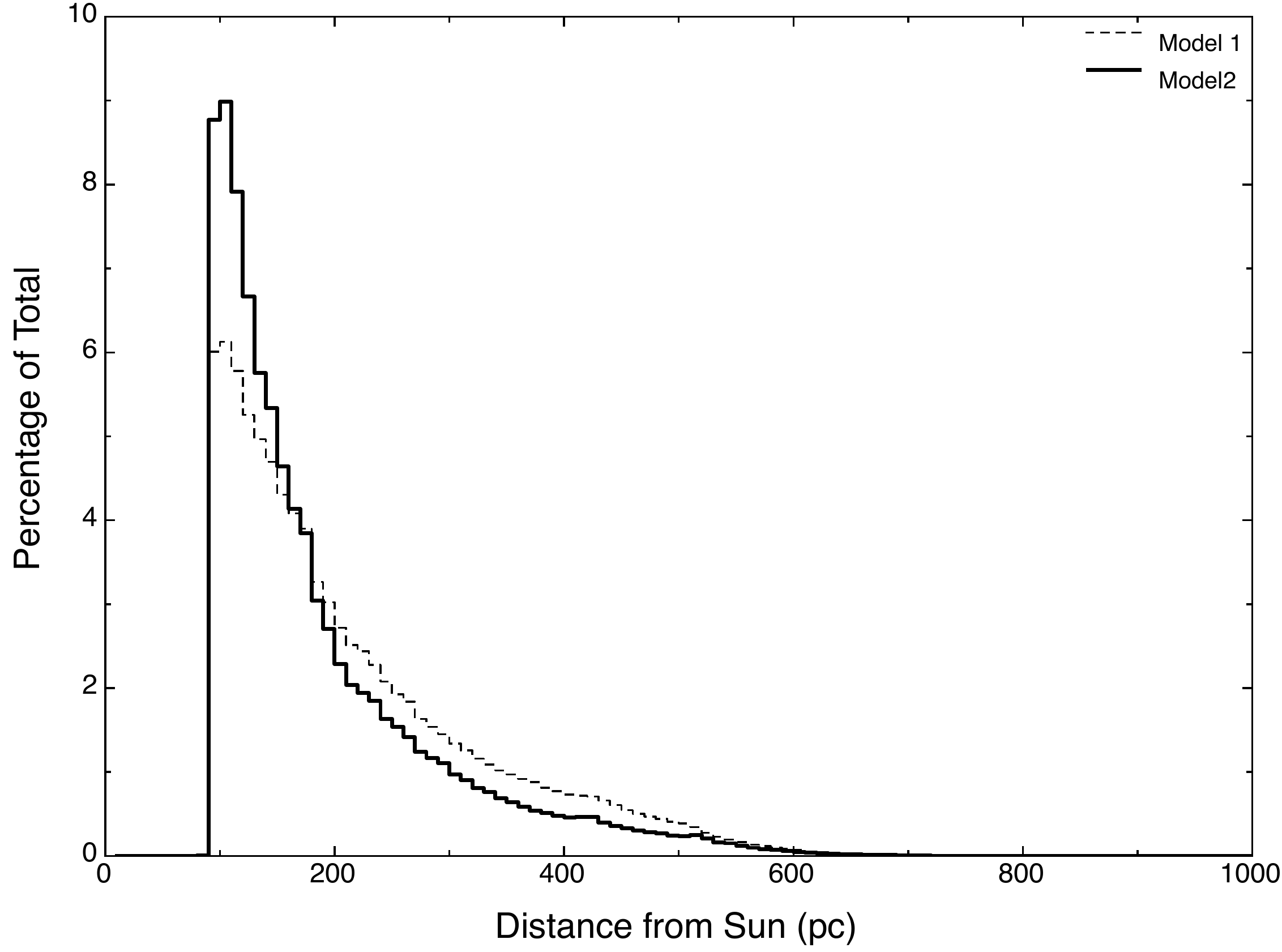}
\caption{The percentage of the flux in each 10 pc bin is shown as a function of distance from the Sun for $a = 0.4$ and $g = 0$ for the two models. Both models include a cavity of 80 pc radius around the Sun.}
\label{flux_with_distance}
\end{minipage}
\hfill
\begin{minipage}[t]{0.4\textwidth}
\includegraphics[width=3.3in]{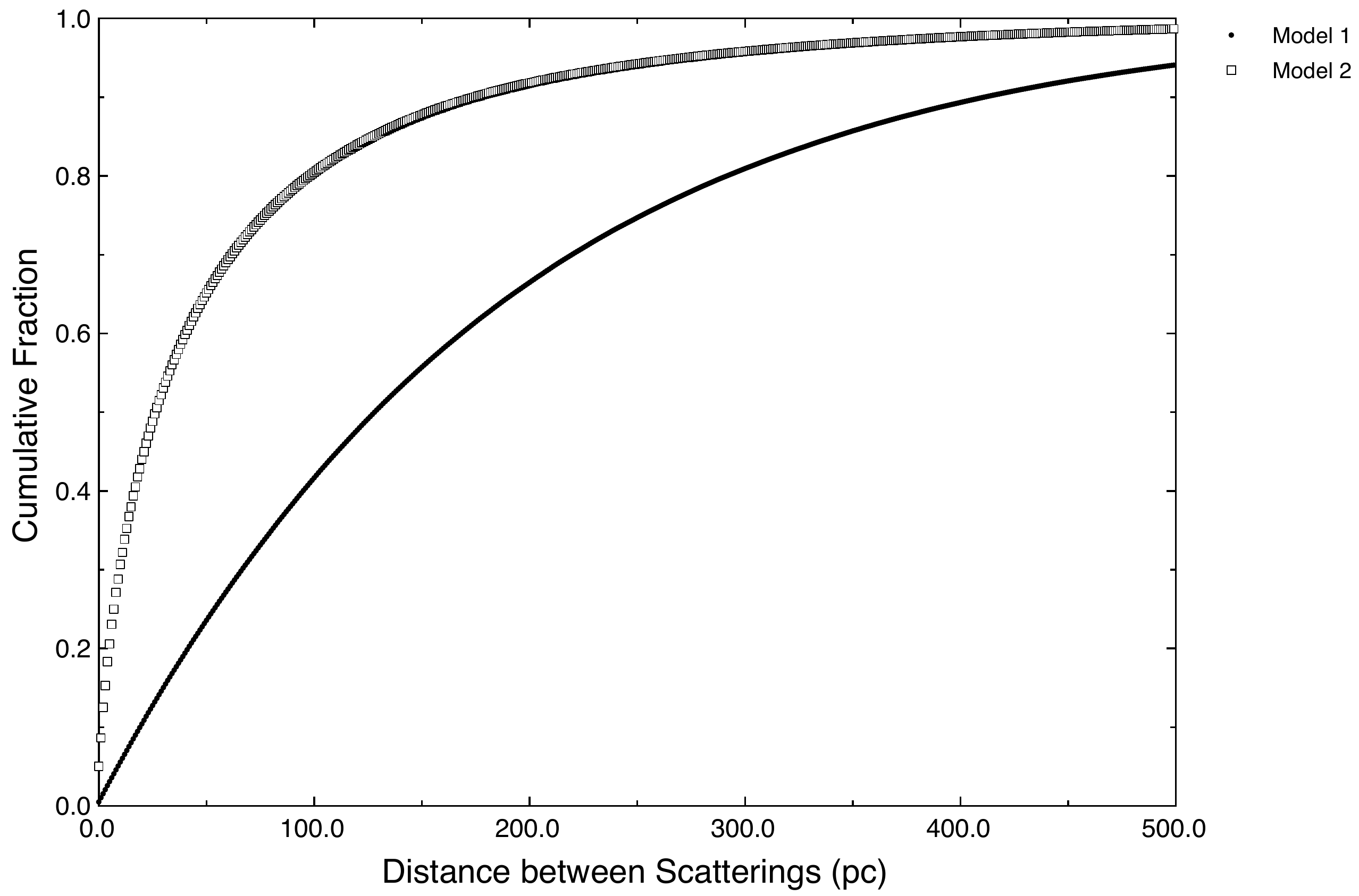}
\caption{The percentage of the total diffuse flux as a function of distance between scatterings is shown.}
\label{distance_between_scattering}
\end{minipage}
\end{figure}

\begin{figure}
\includegraphics[width=4in]{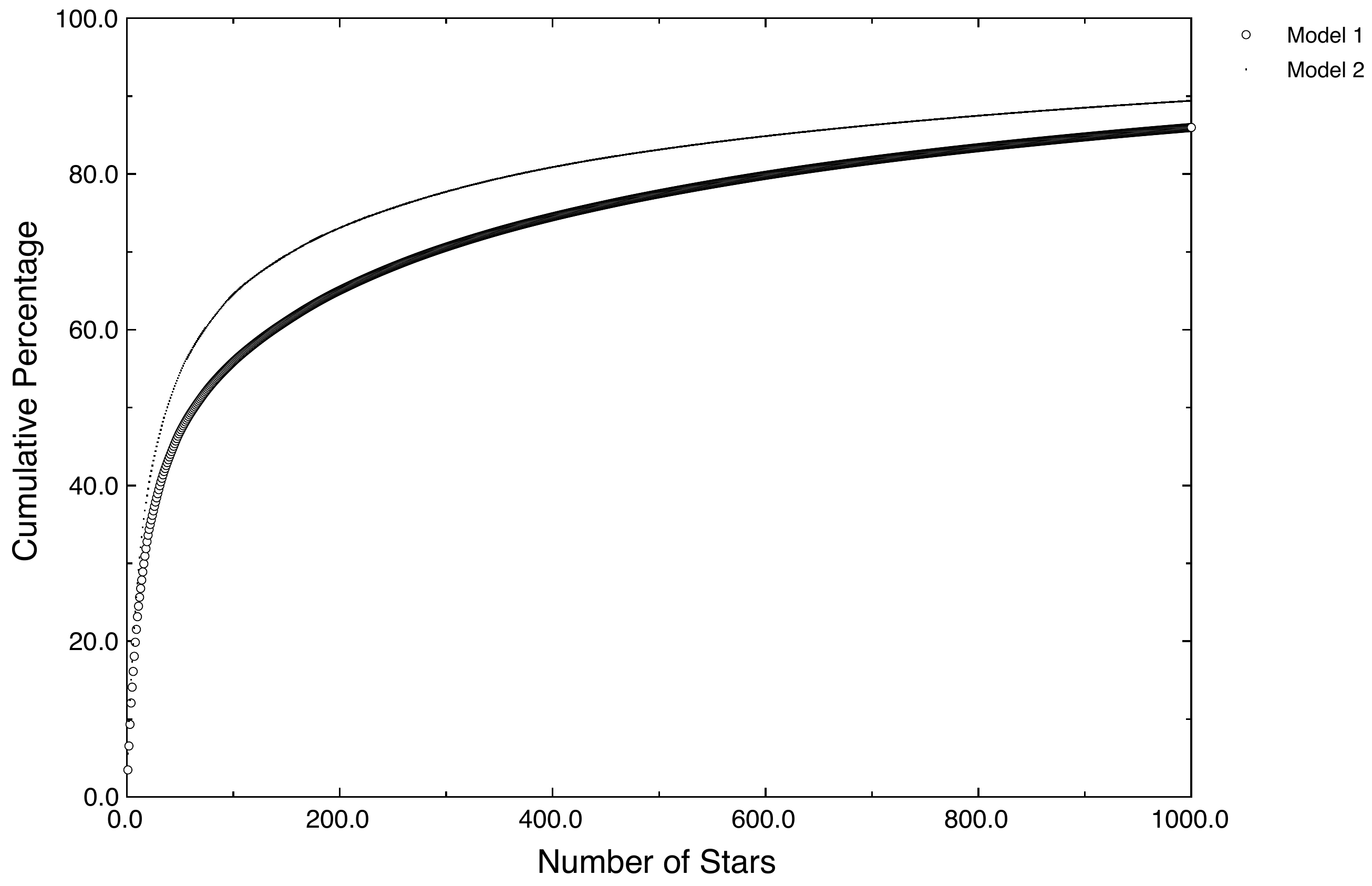}
\caption{The cumulative contribution of the brightest stars is shown for Models 1 and 2.}
\label{cumulative_star}
\end{figure}

The above results depend little on the distribution of the dust. On the other hand the distance of the scattering locations from the Sun depends on the dust distribution but not on the optical parameters. I have plotted the percentage of flux originating in 10 pc bins as a function of distance from the Sun in Fig. \ref{flux_with_distance}. About half of the observed radiation originates within 100 - 200 pc from the Sun and none from cells more than 600 pc away. The average distance between scatters is also dependent on the dust distribution (Fig. \ref{distance_between_scattering}). About 60\% of the photons are scattered within 200 pc of their origin in Model 1 and more than 90\% in Model 2. Virtually all the photons travel less than 500 pc before being scattered. The optical depth per bin is higher in Model 2 (Fig. \ref{pred_model_ebv}) and most photons will not travel more than one optical depth before interacting with a dust grain.

\begin{table}
\caption{Five Brightest Stars}
\label{tab_cumulative_star}
\begin{tabular}{c c c}
\hline
Star Name& HIP No. & Percentage of Total Flux\\
\hline
\multicolumn{3}{c}{Model 1}\\
$\beta$ Cen	    & 68702	& 3.5 \\
$\zeta$ Ori	    & 26727	& 3.1 \\
$\epsilon$ Ori	& 26311	& 2.8 \\
$\alpha$ Cru	& 60718	& 2.7 \\
$\zeta$ Pup	    & 39429	& 2.0 \\
\multicolumn{3}{c}{Model 2}\\
$\alpha$ Cru    & 60718 & 5.6 \\
$\beta$ Cru      & 62434 & 4.2 \\
$\zeta$ Ori     & 26727 & 2.7 \\
$\zeta$ Oph     & 81377 & 2.6 \\
$\epsilon$ Ori  & 26311 & 2.3 \\
\hline
\end{tabular}  
\end{table}

\begin{figure}
\includegraphics[width=4.5in]{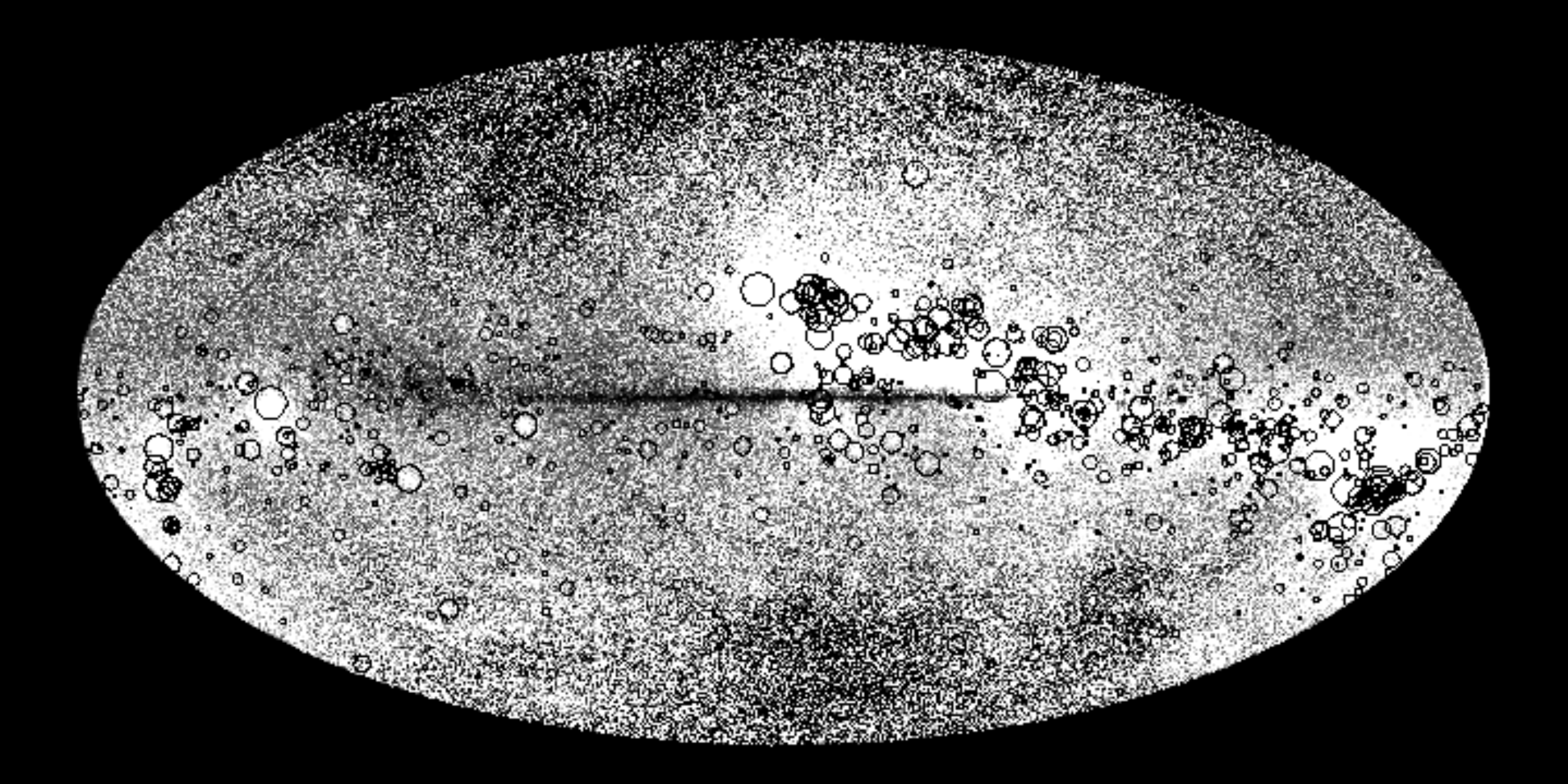}
\includegraphics[width=4.5in]{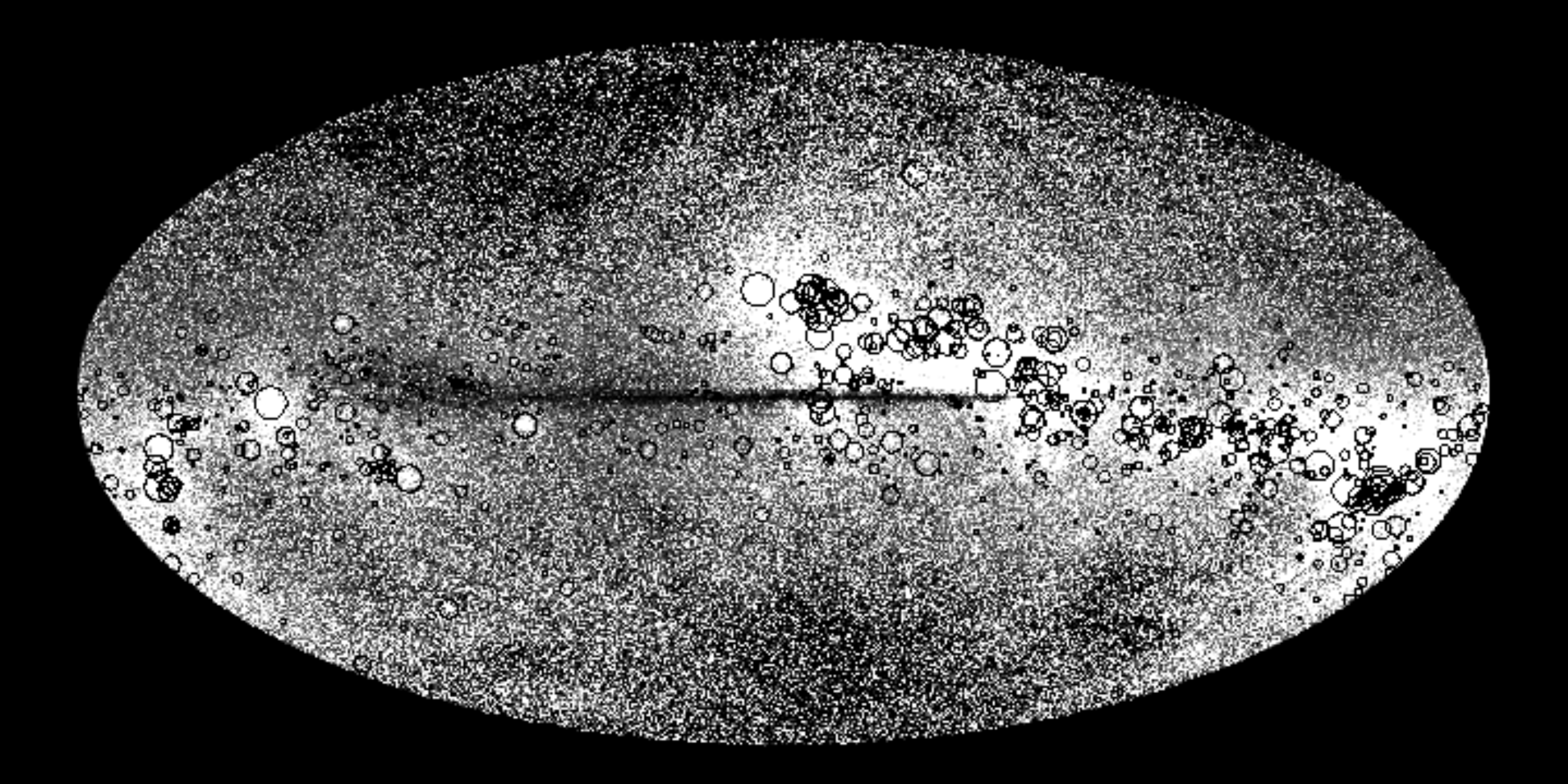}
\includegraphics[width=4.5in]{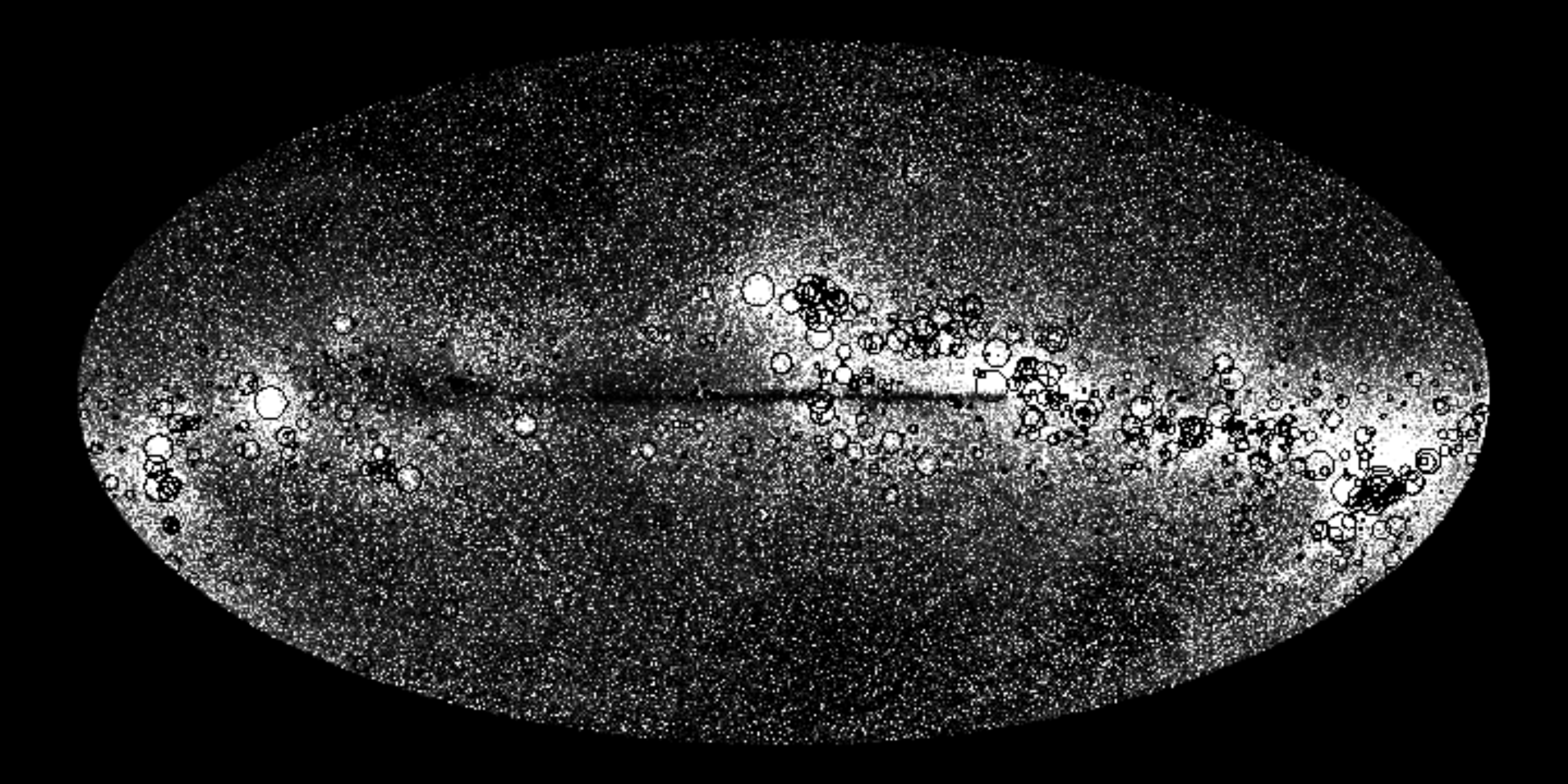}
\caption{Model 2 predictions at 1500 \AA\ for $a = 0.4$ and $g = 0$ (top), $g = 0.4$ (middle), and $g = 0.8$ (bottom). The brightest stars are plotted as circles with the radius proportional to the log of the brightness at 1500 \AA. Similar results are obtained at 2300 \AA\ and for Model 1.}
\label{fuv_model_stars}
\end{figure}

The diffuse flux from our Galaxy is dominated by a handful of stars (Table \ref{tab_cumulative_star}) with 23\% and 27\% of the total flux coming from only 10 stars for Models 1 and 2, respectively, and  90\% of the total observed flux (Fig. \ref{cumulative_star}) from the 1000 brightest stars. I have plotted the predictions from Model 2 at 1500 \AA\ in Fig. \ref{fuv_model_stars} for $a = 0.4$ and $g = 0$ (isotropic scattering), $g = 0.5$ and $g = 0.8$. The diffuse light is concentrated in Gould's Belt following the stars, plotted as circles with a radius proportional to the log of the star brightness. As the grains become more forward scattering, the diffuse light becomes more localized near the stars.

\begin{figure}
\includegraphics[width=3.2in]{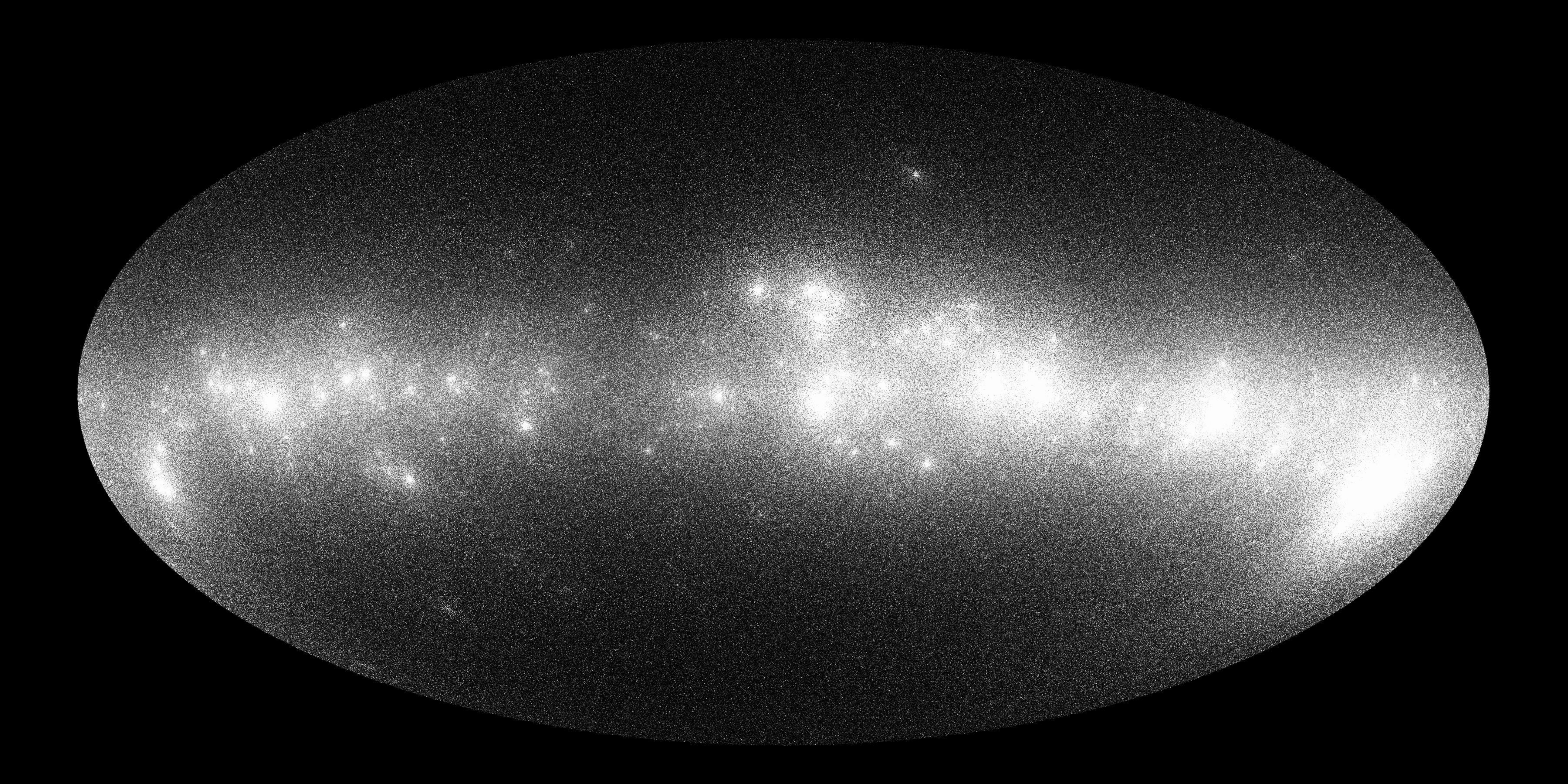}
\includegraphics[width=3.2in]{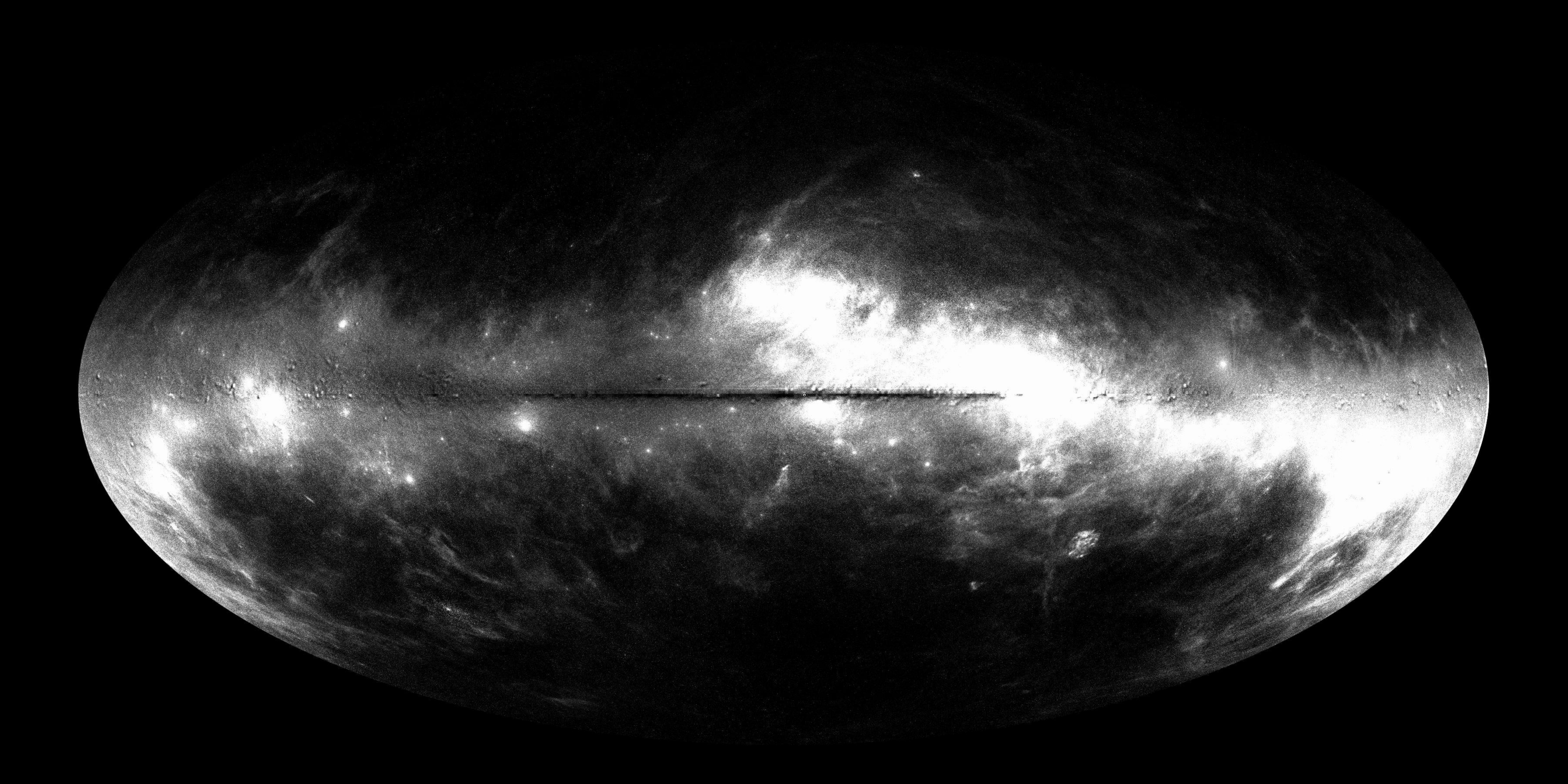}
\caption{Scattered light predictions for Model 1 (top) and Model 2 (bottom) at 1500 \AA\ for $a = 0.36$ and $g = 0.5$.}
\label{fuv_model_comparison}
\end{figure}

\begin{figure}
\includegraphics[width=3in]{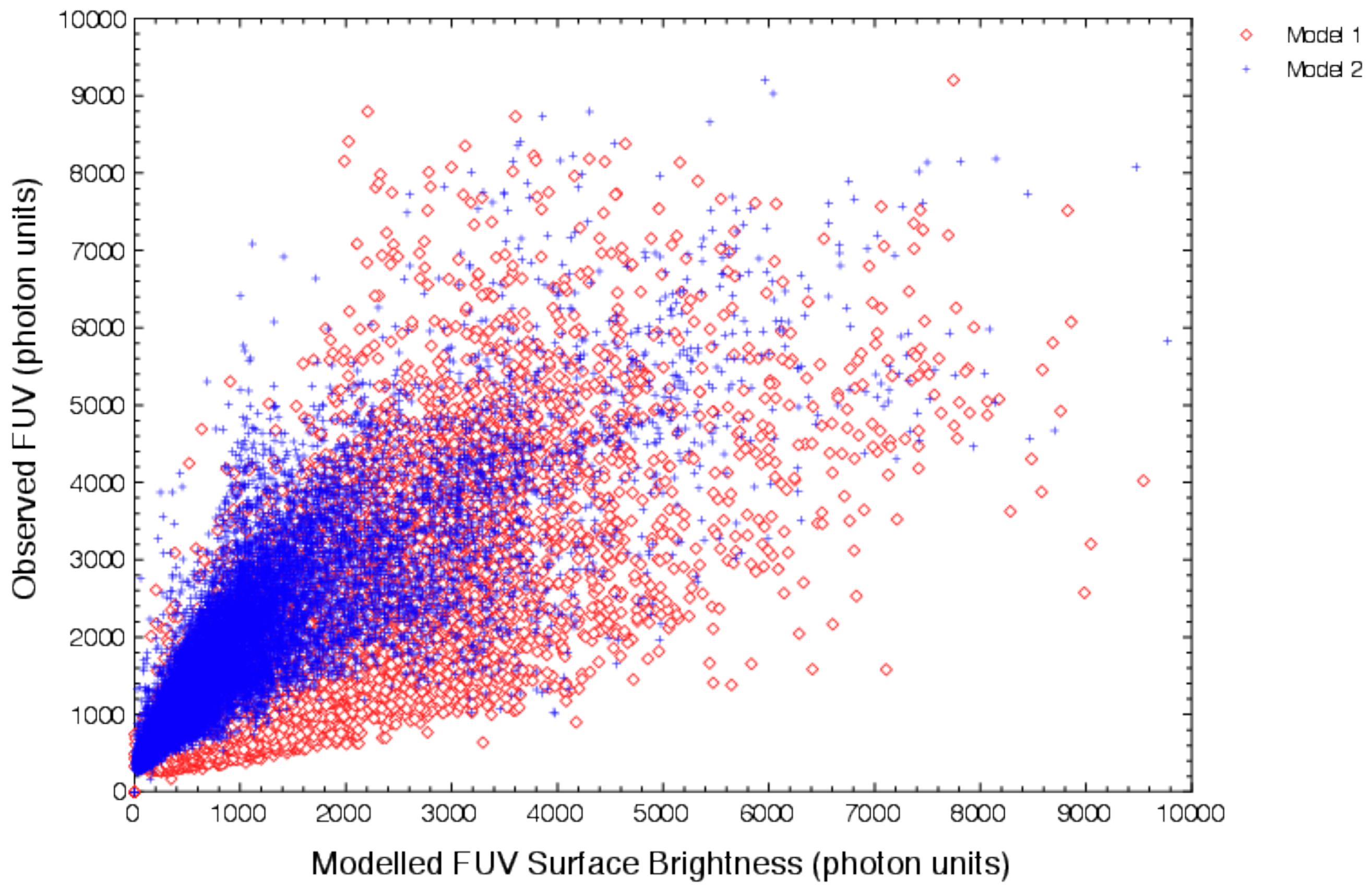}
\includegraphics[width=3in]{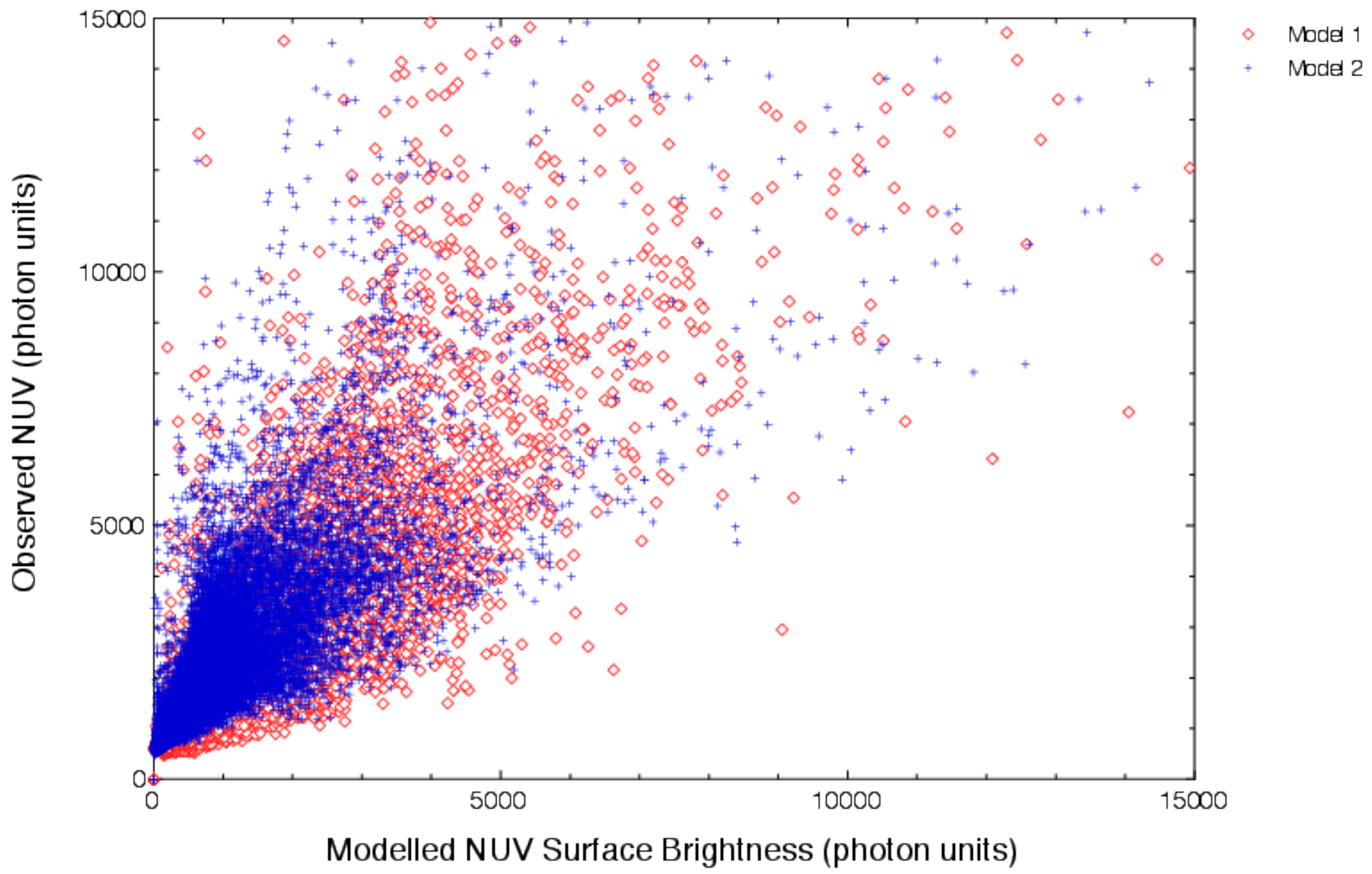}
\caption{FUV (top) and NUV (bottom) plotted versus model predictions for Model 1 (red) and Model 2 (blue).}
\label{observed_modeled_flux}
\end{figure}

Also superficially similar are the calculated diffuse backgrounds over the entire sky for each of two dust distributions (Fig.  \ref{fuv_model_comparison}) with both showing a good correlation with the data (Fig. \ref{observed_modeled_flux}). The correlation between the two models is best at low latitudes (FUV: r = 0.841; NUV: r= 0.808) where the optical depth is high and the scattered light is concentrated near the stars and is poor at high latitudes (FUV: r = 0.441; NUV: r = 0.360) where the optical depth is low and the scattered light is dependent on the dust distribution. The total output from the Galaxy is dominated by the low latitude dust and hence will not be sensitive to the details of the dust distribution in the Galaxy.

\section{Comparison with Data}

\begin{table}
\caption{Best Fit Parameters}
\label{best_fit_opt_consts}
\begin{tabular}{l c c c c c}
\hline
Model & a & g & y$_{0}$ & $\chi^{2}$ & r\\
\hline
\multicolumn{6}{c}{FUV}\\
Model 1 (all $\tau$) & 0.3  & 0.0 & -68 &	3.91 & 0.750\\
Model 1 ($\tau < 1$) & 0.2  & 0.1 & 253 & 2.75   & 0.721\\
Model 1 ($\tau > 1$) & 0.25 & 0.0 & 978 & 14.96  & 0.591\\
Model 2 (all $\tau$) & 0.36 & 0.0 & 391 & 2.99   & 0.861\\
Model 2 ($\tau < 1$) & 0.50 & 0.3 & 224 & 1.91   & 0.888\\
Model 2 ($\tau > 1$) & 0.30 & 0.1 & 1026 & 13.64 & 0.681\\
\multicolumn{6}{c}{NUV}\\
Model 1 (all $\tau$) & 0.40  & 0.3 & 356 &	4.48 & 0.810\\
Model 1 ($\tau < 1$) & 0.30  & 0.1 & 367 & 2.74   & 0.733\\
Model 1 ($\tau > 1$) & 0.40 & 0.2 & 732 & 14.81  & 0.699\\
Model 2 (all $\tau$) & 0.39 & 0.2 & 859 & 4.89   & 0.769\\
Model 2 ($\tau < 1$) & 0.50 & 0.4 & 627 & 2.10   & 0.857\\
Model 2 ($\tau > 1$) & 0.31 & 0.4 & 1818 & 16.74 & 0.592\\
\hline
\end{tabular}  
\end{table}

\begin{figure}
\includegraphics[width=3in]{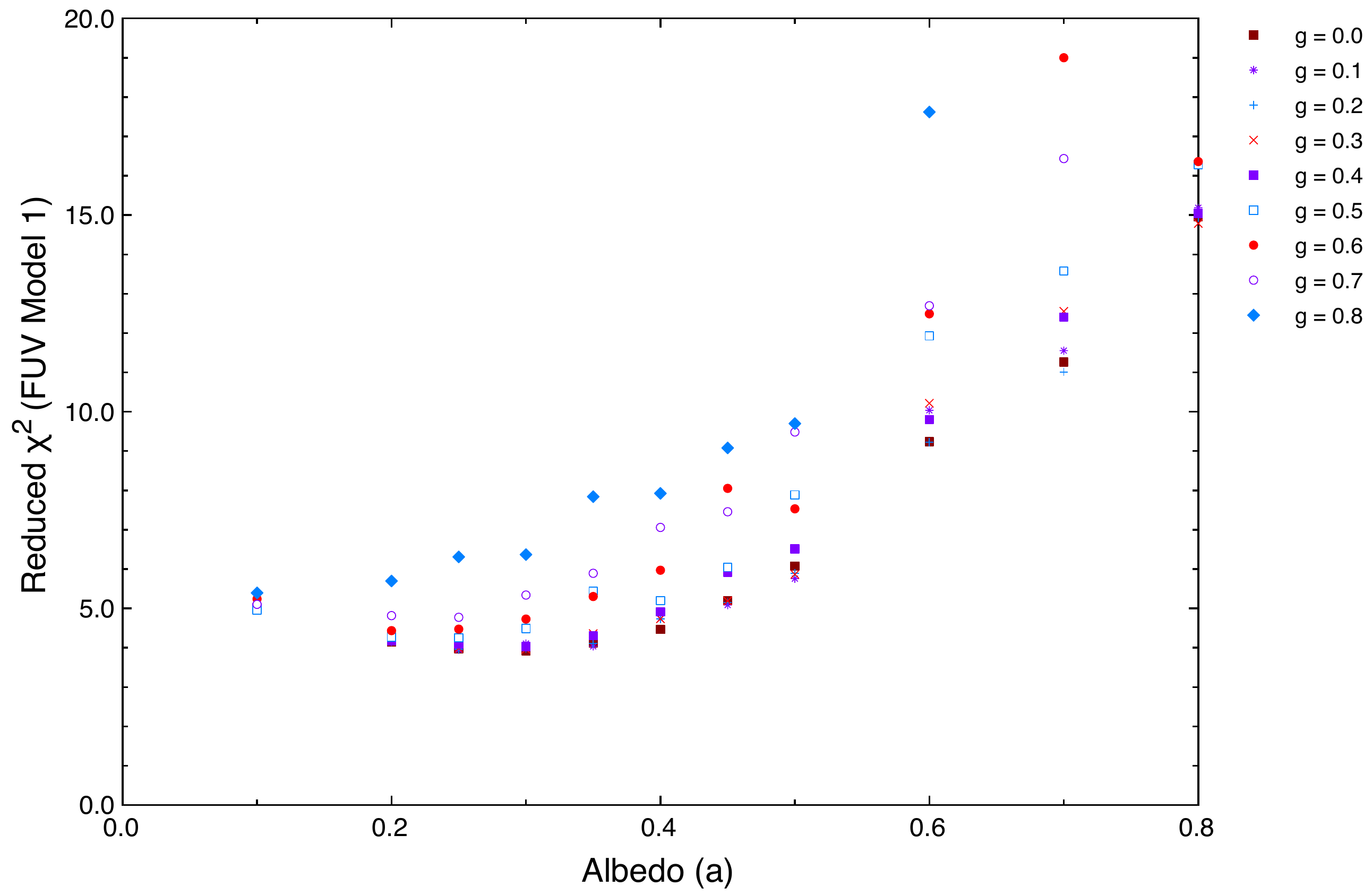}
\includegraphics[width=3in]{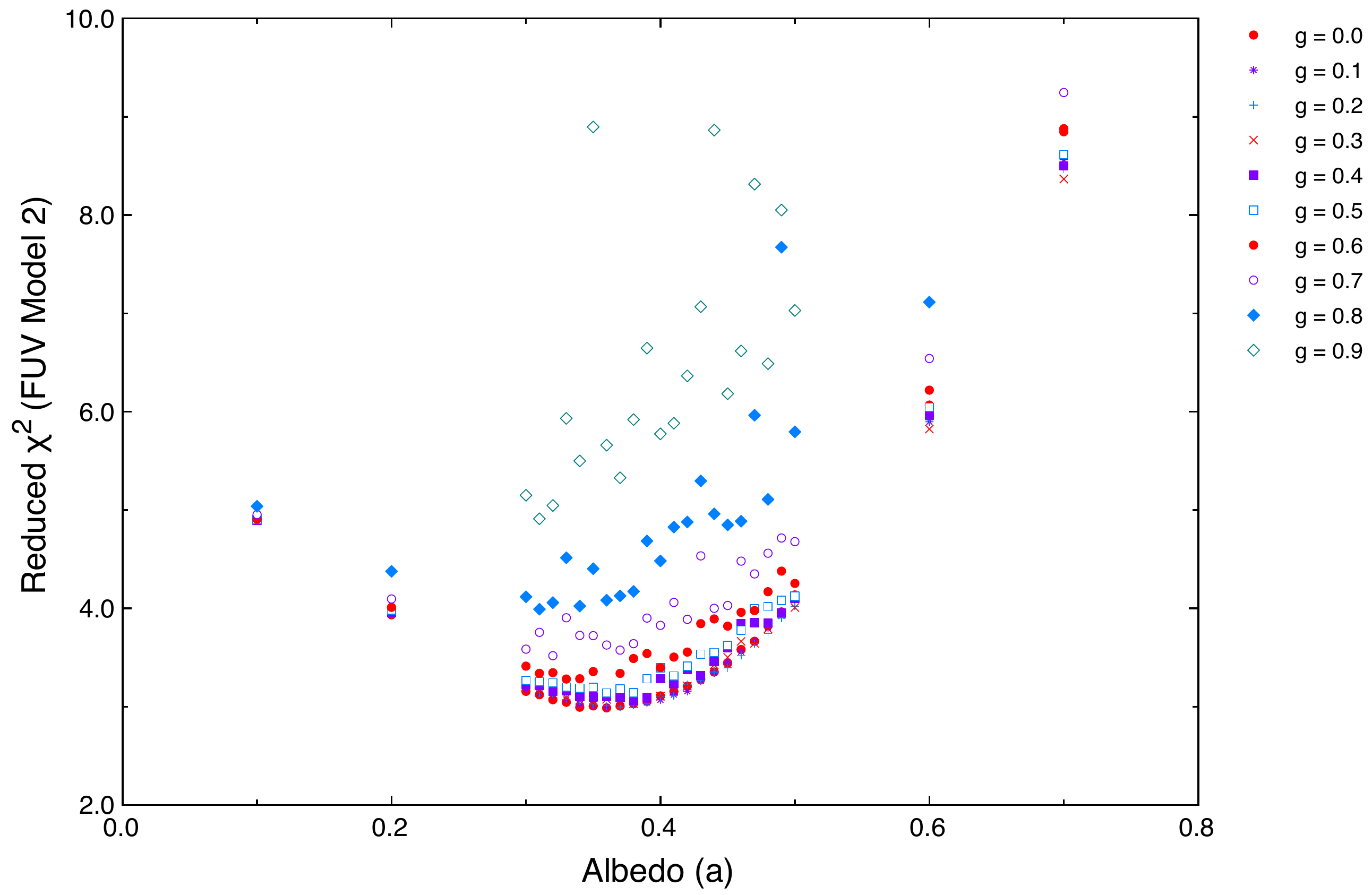}
\caption{Reduced $\chi^{2}$ plotted as a function of the optical constants in the FUV band for Model 1 (top) and Model 2 (bottom).}
\label{fuv_obs_chisq}
\end{figure}

\begin{figure}
\includegraphics[width=3in]{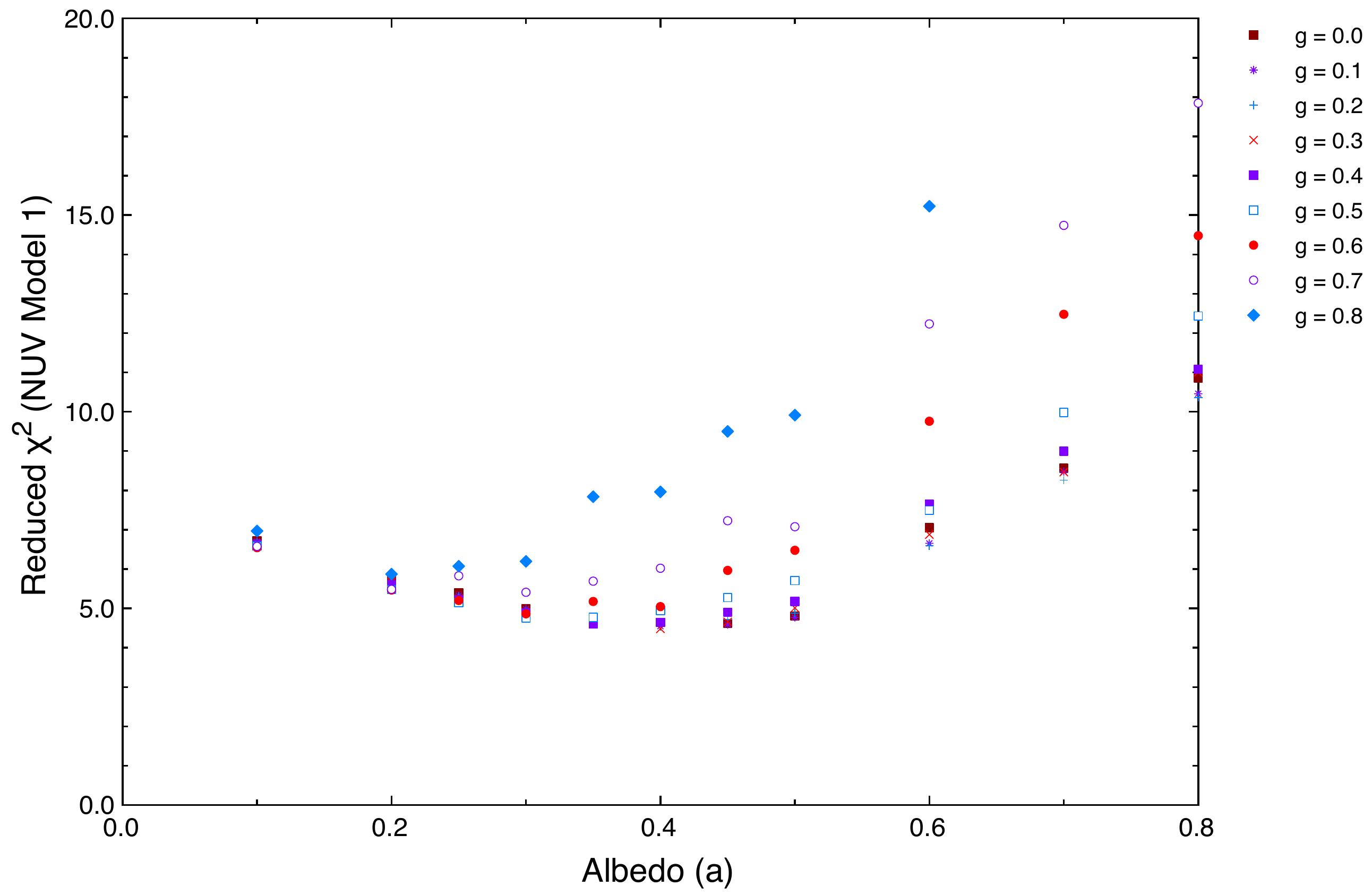}
\includegraphics[width=3in]{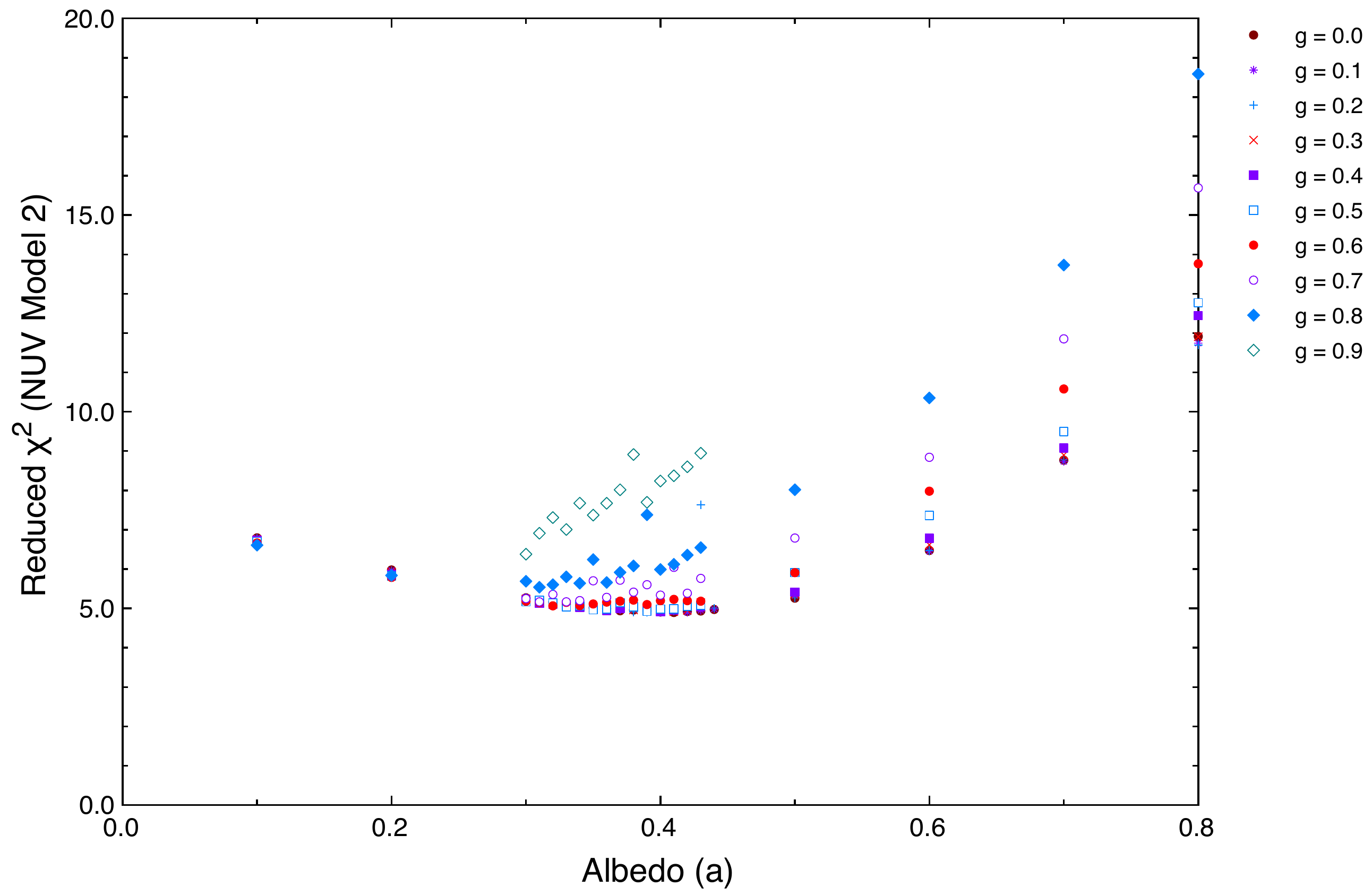}
\caption{Reduced $\chi^{2}$ plotted as a function of the optical constants in the NUV band for Model 1 (top) and Model 2 (bottom).}
\label{nuv_obs_chisq}
\end{figure}

The main motivation for this study was the availability of the all-sky \galex\ data which allows tests of the scattering on both small and large scales. I have compared the output of the Monte Carlo models for both dust distributions for a range of optical constants in the FUV band (Fig. \ref{fuv_obs_chisq}) and in the NUV (Fig. \ref{nuv_obs_chisq}). In each case, there was an offset between the model and the observations representing components of the diffuse background other than dust scattered light and I added these offsets to the model output before calculating the reduced $\chi^{2}$ of the fit between the models and the data (Table 2). These tests were carried out assuming a $500 \times 500 \times 500$ grid for the Galaxy with each run incorporating $5 \times 10^{8}$ photons.

\begin{table}
\caption{Latitudinal Fits ($a = 0.36; g = 0.5$)}
\label{lat_fit_opt_consts}
\begin{tabular}{l c c c c}
\hline
Range & Slope & y$_{0}$ & $\chi^{2}$ & r\\
\hline
\multicolumn{5}{c}{FUV}\\
$|b| < 30; \tau < 1$ & 1.24 & 741  &  5.03 &  0.846\\
$|b| < 30; \tau > 1$ & 0.72 & 1430 & 13.74 &  0.724\\
$30 < b < 60; \tau < 1$        & 1.79 &  359 & 1.93  &  0.917\\
$-60 < b < -30; \tau < 1$      & 1.01 &  457 & 2.17  &  0.856\\
$60 < b$             & 1.15  & 325 & 1.26  &  0.468\\
$-60 > b $           & 0.92  & 313 & 1.09  &  0.592\\

\multicolumn{5}{c}{NUV}\\
$|b| < 30; \tau < 1$ & 1.57 &  895  &  5.75  & 0.822\\
$|b| < 30; \tau > 1$ & 0.63 &  2241 & 17.43  & 0.589\\
$30 < b < 60; \tau < 1$        & 1.85 &  592 &  1.53  & 0.890\\
$-60 < b < -30; \tau < 1$      & 1.12 &  717  & 3.52  & 0.592\\
$60 < b$             & 1.20 &  590  & 1.12  & 0.437\\
$-60 > b $           & 0.54  & 637  & 1.29  & 0.266\\
\hline
\end{tabular}  
\end{table}

The predictions of both models for the dust distribution are generally consistent with the data suggesting that, at least on a global scale, an accurate knowledge of the dust distribution may not be critical in determining the diffuse background. Rather, much of the structure seen in the diffuse background is due to the stellar distribution. This is even more apparent at still shorter wavelengths \citep{Murthy2004} where there are many fewer bright stars.

I will look more closely at the distribution of the background in different sections of the Galaxy in the following paragraphs but will focus on Model 2, where the dust distribution is more closely reproduced. The signal-to-noise of the Monte Carlo runs is a limiting factor at high latitudes and I have used a single long run of $8 \times 10^{9}$ photons with fixed optical constants of ($a = 0.36; g = 0.5$). These values fall near the $\chi^{2}$ minimum and are consistent with other determinations in the literature (reviewed by \citet{Draine2003}). I have found the best fit of the model to the data in the different latitude intervals and tabulated these in Table \ref{lat_fit_opt_consts}. I have allowed for a slope and an offset between the model and the data where the slope may indicate either differences in the albedo from $a = 0.36$ or an additional component with the same distribution as the scattered light. The offset represents additional contributors to the diffuse background perhaps including an airglow component and will be discussed further below. 

\begin{figure}
\includegraphics[width=3in]{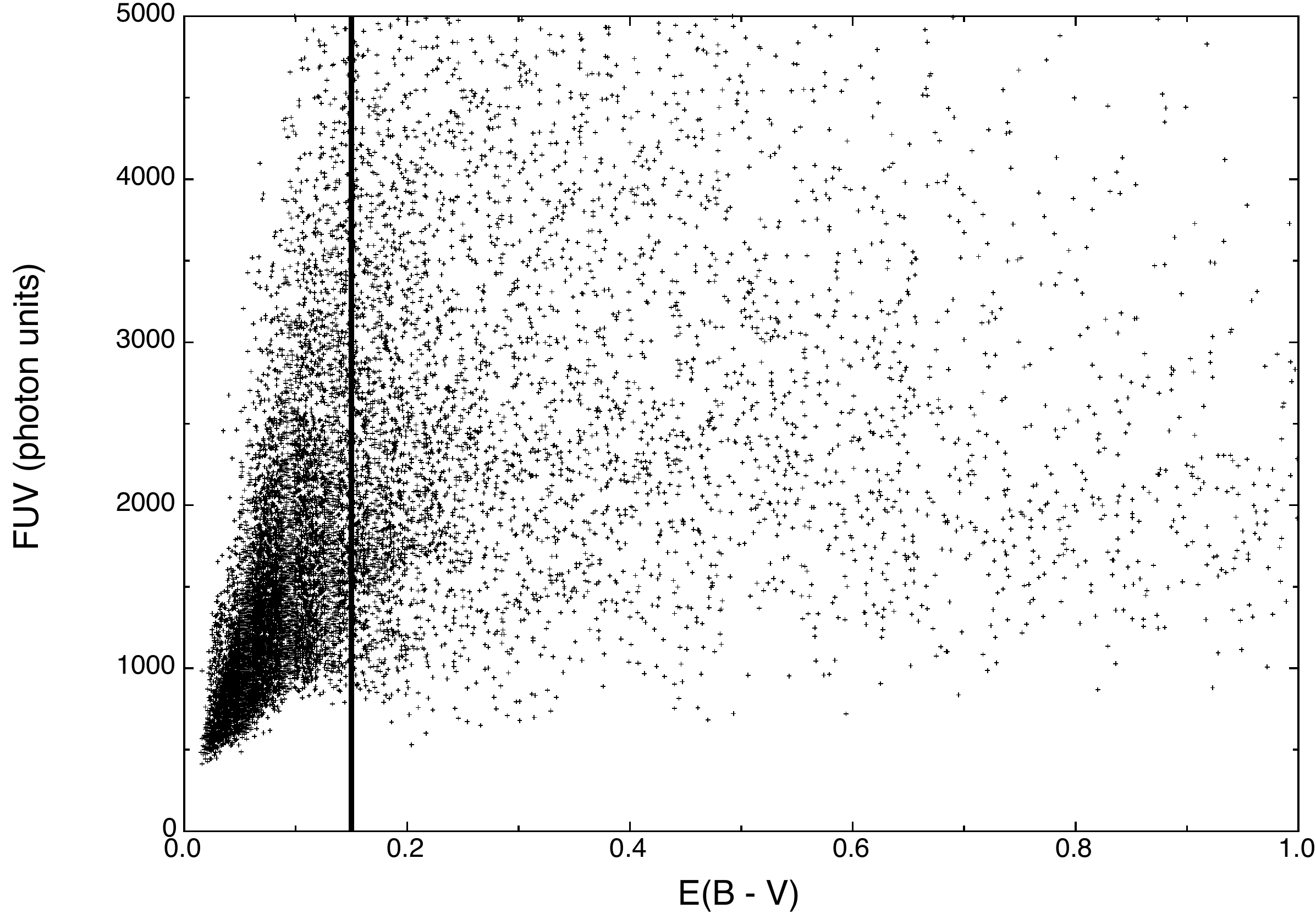}
\caption{Observed FUV flux plotted versus E(B - V) for $|b| < 30^{\circ}$. The solid line represents an E(B - V) of 0.15 which corresponds to an optical depth of 1 in both the FUV and NUV \galex\ bands. The NUV plot is similar and I have not shown it.}
\label{ebv_fuv_low_latitude}
\end{figure}

\begin{figure}
\includegraphics[width=3in]{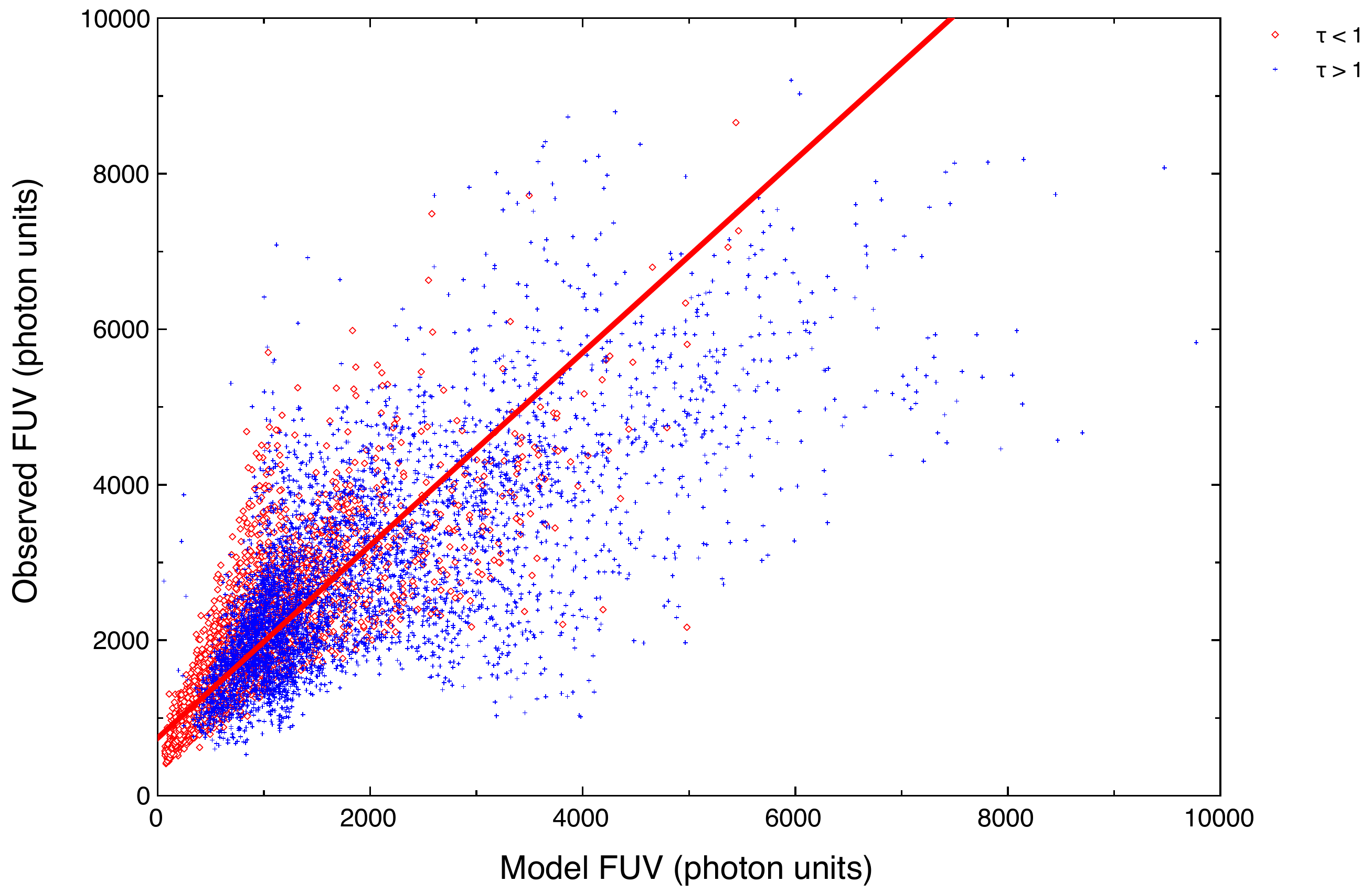}
\includegraphics[width=3in]{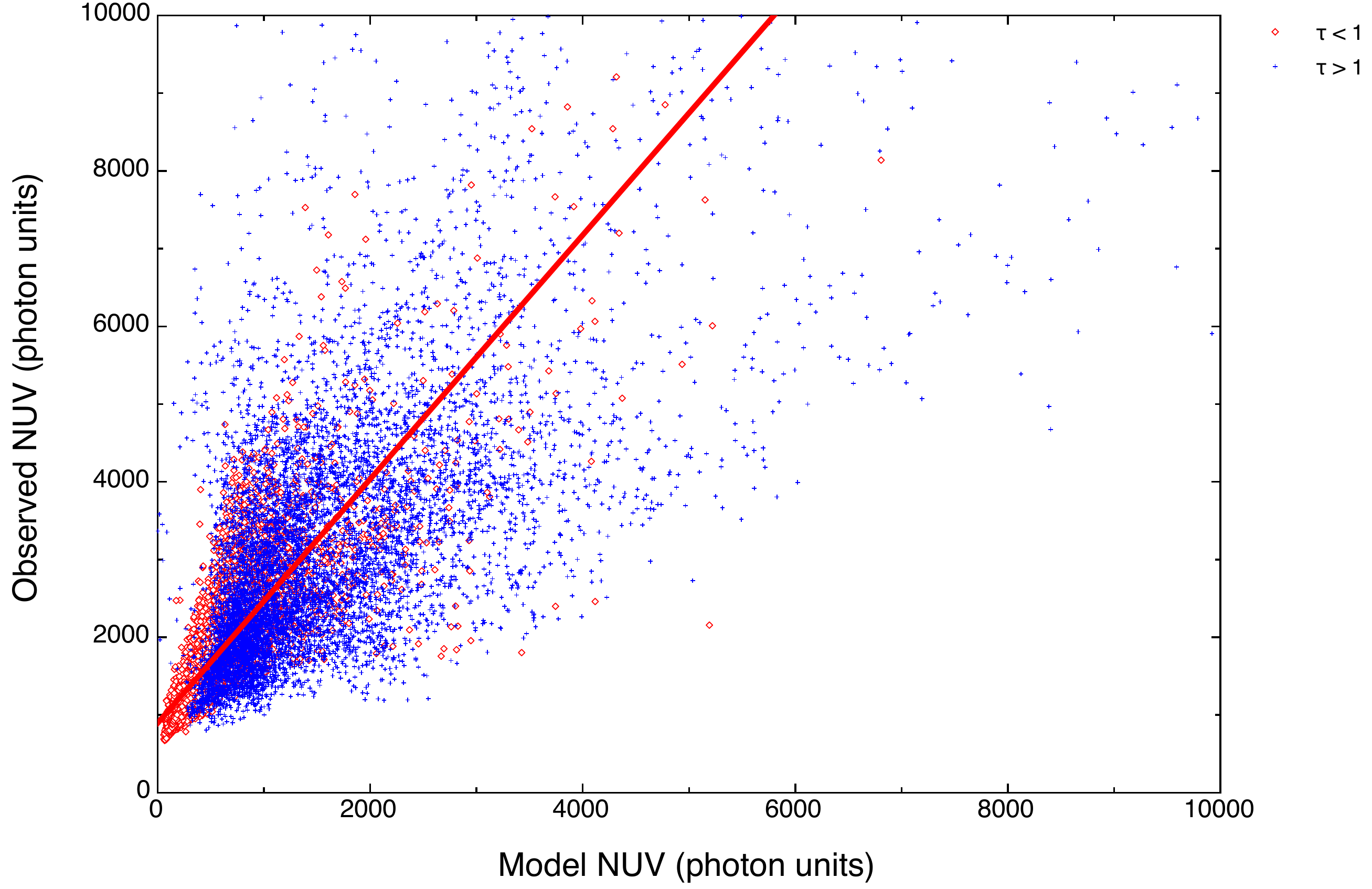}
\caption{Observed flux at low Galactic latitudes plotted versus predicted flux for $\tau < 1$ (red) and $\tau > 1$ (blue) for FUV (top) and NUV (bottom) bands. The red line represents the best fit between the model and the data. I have not plotted the best fit line for $\tau > 1$ because of the scatter in the data.}
\label{low_lat_models}
\end{figure}

\subsection{Mid and Low Latitudes}

The range of E(B - V) is greatest at low latitudes with the expected areas of high extinction near the plane but with surprisingly low values even short distances away from the plane. I have plotted the observed FUV as a function of E(B - V) in Fig. \ref{ebv_fuv_low_latitude} and divided the observations into two regions: $E(B - V) < 0.15$ and $E(B - V) > 0.15$, approximately corresponding to an optical depth of 1 in both \galex\ bands. There is a correlation between the observed background and the reddening for $\tau < 1$ (FUV: r = 0.583; NUV: r = 0.489) and the model does a good job of predicting the background with correlations of 0.846 and 0.822 in the FUV and NUV (Fig. 16). The better correlations of the model to the data are because the diffuse background is dependent on the distribution of the dust and of the stars, which is accounted for by the modelling.

The slope in both bands is somewhat greater than 1 suggesting that the albedo was underestimated by a factor equal to the square root of the slope (Fig. \ref{flux_with_albedo}) implying that $a = 0.4$ in the FUV and $a = 0.45$ in the NUV. There is an offset of 740 and 900 \photu\ in the FUV and the NUV bands, respectively, of which \citet{Murthy2014a} attributed 200 --- 300 \photu\ to unresolved airglow. Other contributors to the diffuse background at low latitudes may include molecular hydrogen fluorescence in the FUV \citep{Hurwitz1994, Lim2013} or other more exotic sources \citep{Henry2015}. I have not included these sources in my model but will do so in a future version.

There is considerable scatter between the observations and the reddening for $\tau > 1$ (FUV: r = -0.001; NUV: r = 0.120) because only the front layers of the dust contribute to the observed background (Fig. \ref{ebv_fuv_low_latitude}). The models fit the data reasonably well  (FUV: r = 0.724; NUV: r = 0.589) but with considerable scatter although the models should include self-extinction by the dust. This is because the distribution of dust in these line of sight is likely to be more complex and local effects may determine the observed background. One example of this was seen in the vicinity of the Coalsack Nebula where \citet{Murthy1994} found that the intense diffuse emission was due to the scattering of the light of only three bright stars by a thin layer of dust in front of the dense molecular cloud.

\begin{figure}
\includegraphics[width=3in]{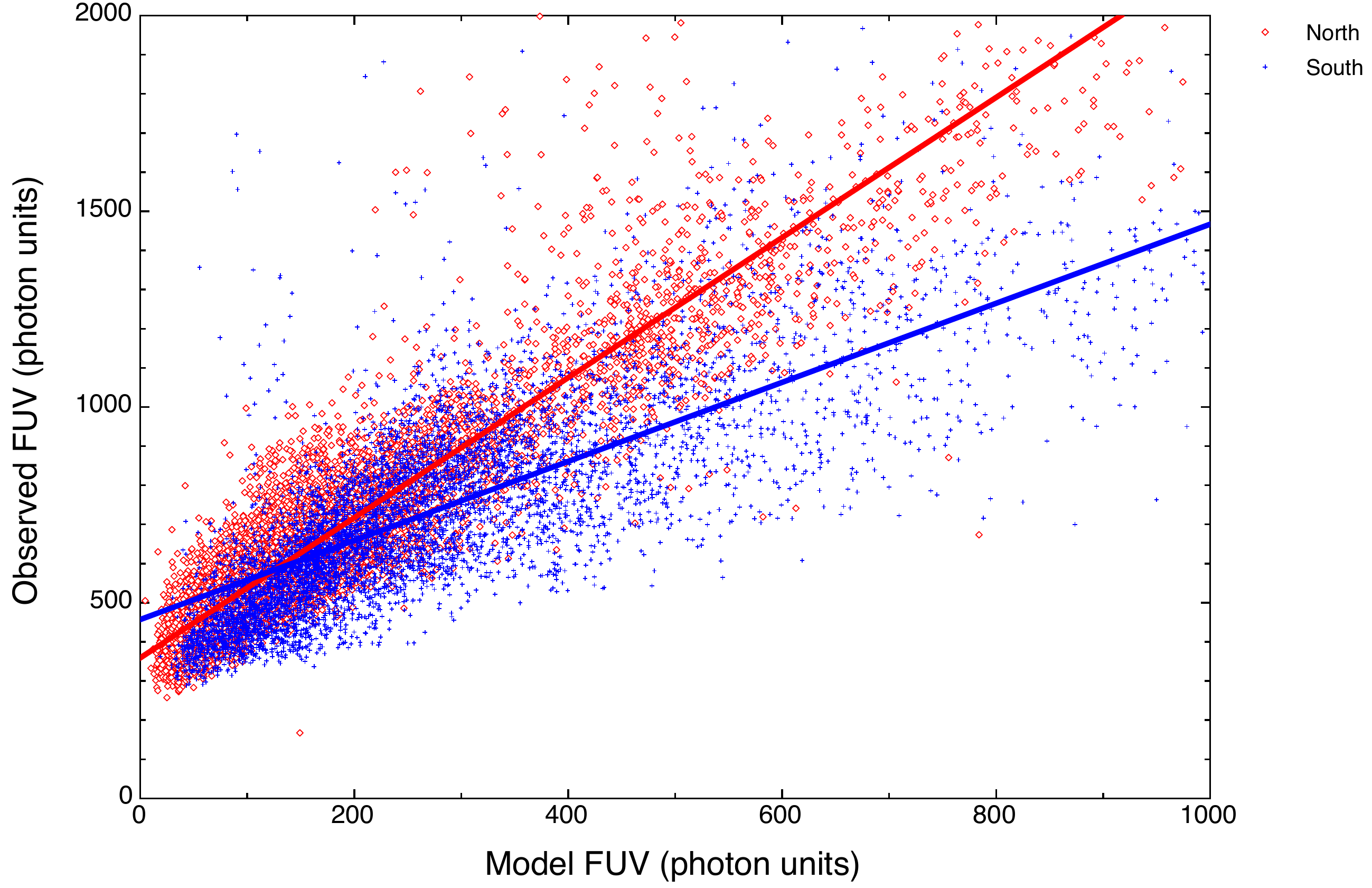}
\includegraphics[width=3in]{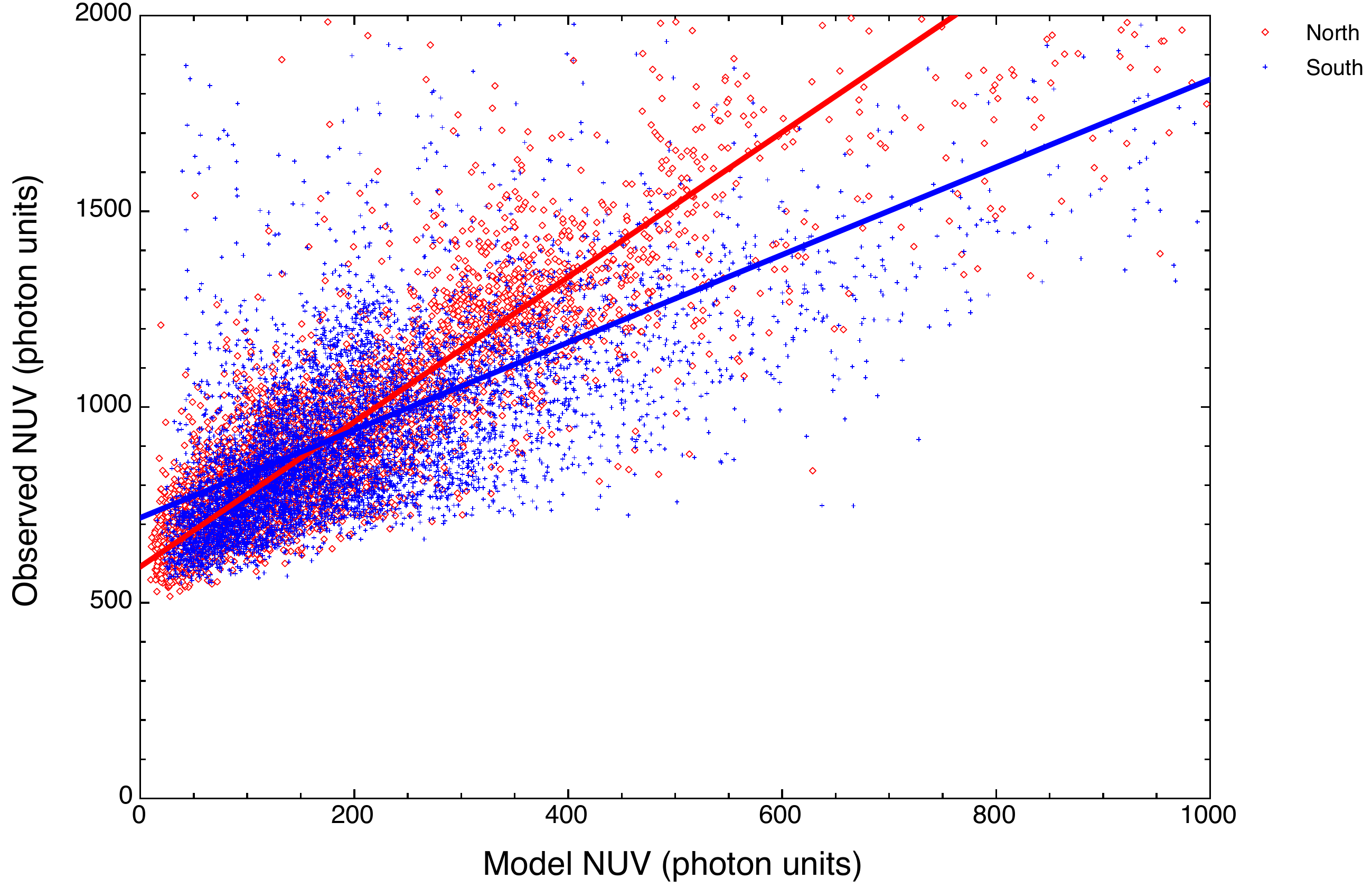}
\caption{Observed flux at mid-latitudes plotted versus predicted flux for FUV (top) and NUV (bottom). The Northern hemisphere points are plotted in blue and the Southern hemisphere in red. The blue and red lines represent linear fits to both hemispheres.}
\label{mid_lat_models}
\end{figure}

There is much less dispersion in E(B - V) at mid-latitudes ($30 < |b| < 60$) and there is a good correlation between the modelled and the observed fluxes (Fig. 17) in both the FUV and the NUV (Table \ref{lat_fit_opt_consts}) with an offset of 350 --- 450  and 600 --- 700 \photu\ in the FUV and NUV bands. Although there is a difference in the slope of the best fit lines between the Northern and the Southern hemispheres in both bands, this is largely because the model under-predicts the brightest points in the Northern hemisphere. The reason for this is unclear and I will defer an explanation pending further modelling.

\begin{figure}
\includegraphics[width=3in]{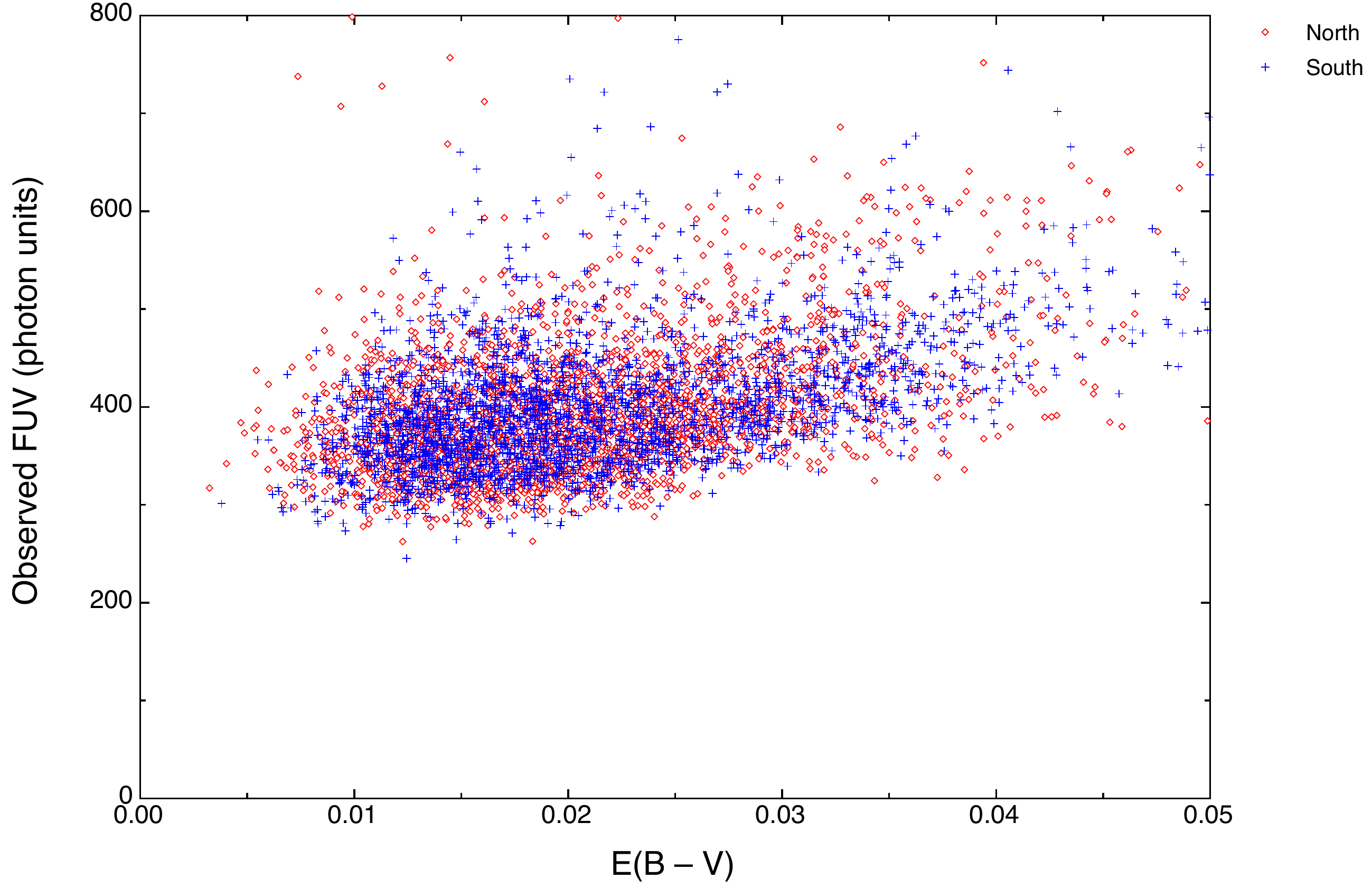}
\includegraphics[width=3in]{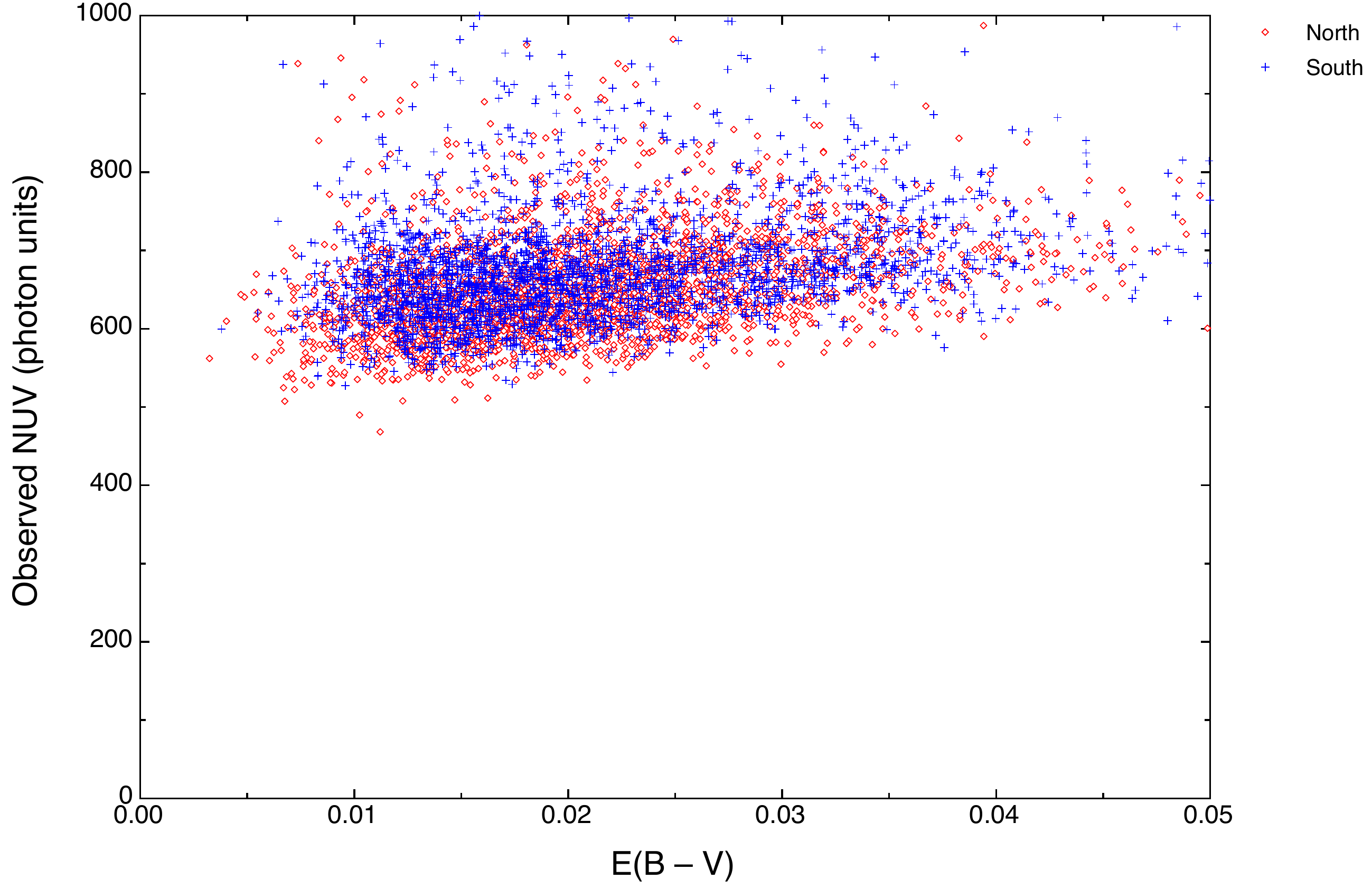}
\caption{Observed flux for FUV (top) and NUV (bottom) at high latitudes plotted versus the reddening. The red points are Southern hemisphere and the blue points are Northern hemisphere.}
\label{ebv_high_lat}
\end{figure}

\begin{figure}
\includegraphics[width=3in]{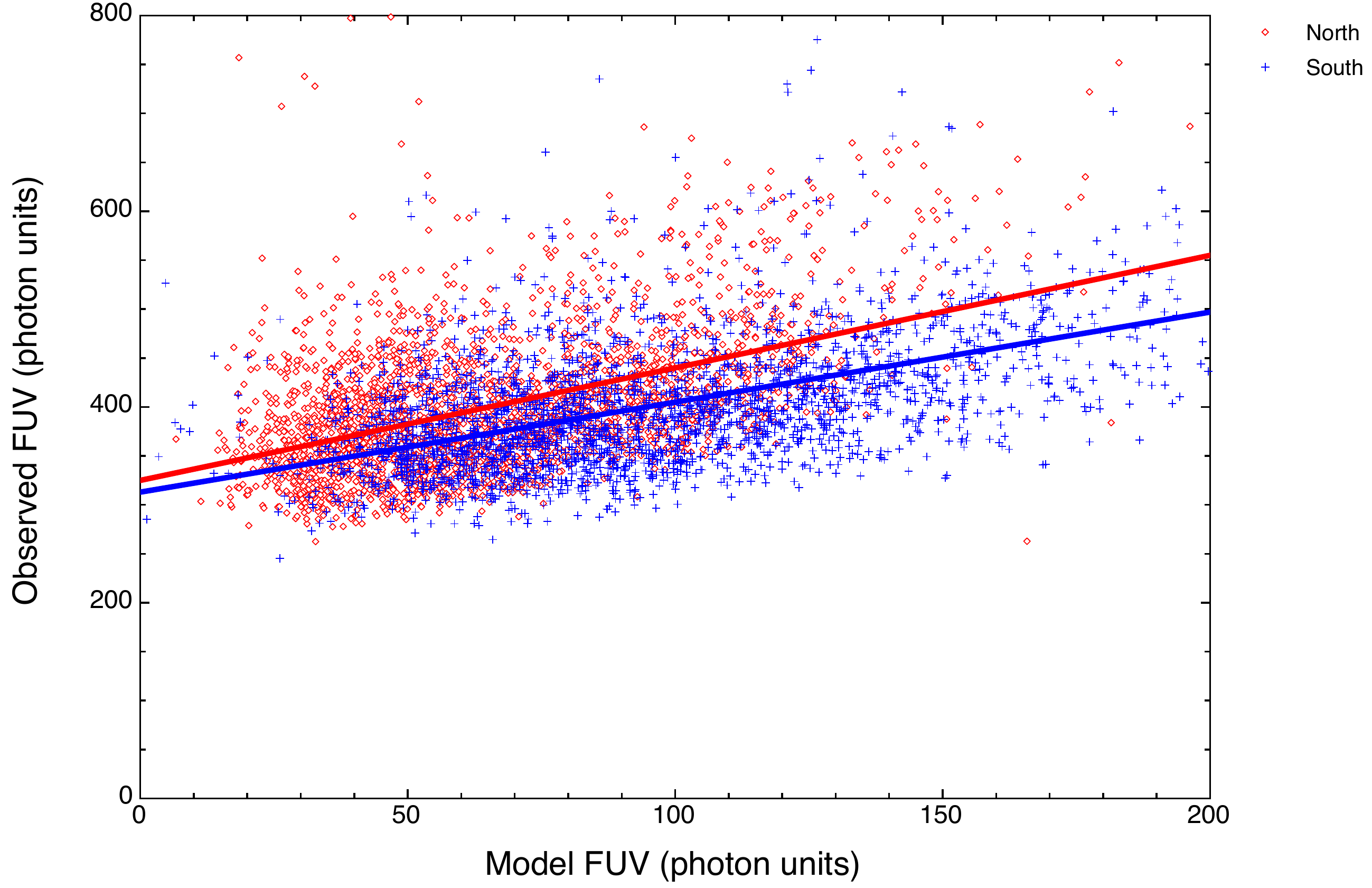}
\includegraphics[width=3in]{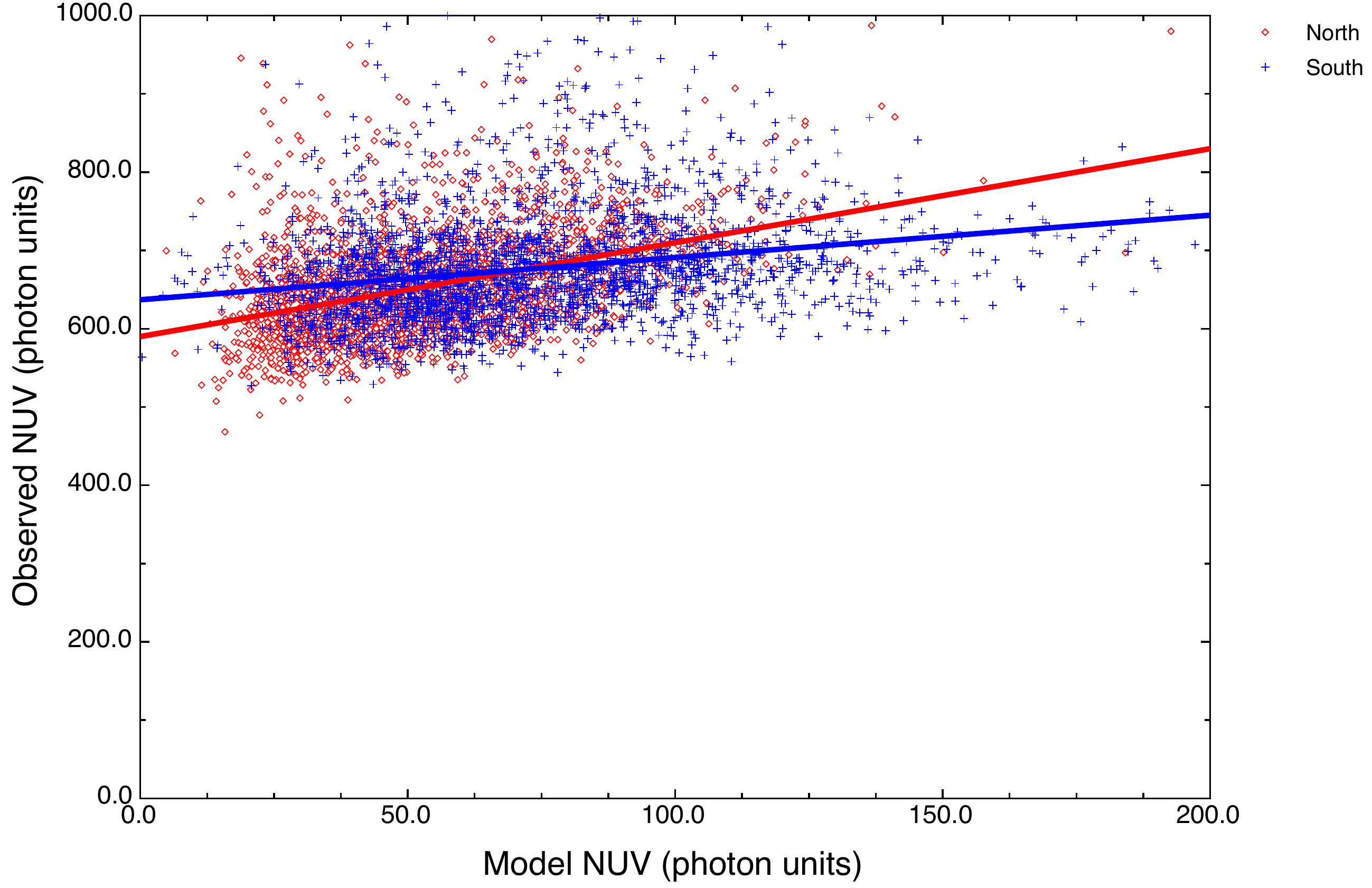}
\caption{Observed flux at high latitudes plotted versus predicted flux. Colours are the same as Fig. \ref{mid_lat_models}.}
\label{high_lat_models}
\end{figure}

\subsection{High Latitudes}

The dust scattered light at high latitudes is due to the back scattering of light from Galactic plane stars \citep{Jura1979} and would be expected to be proportional to the reddening at these low column densities. However, there is considerable scatter in a plot of the observed background as a function of the reddening (Fig. \ref{ebv_high_lat}) perhaps due to uncertainties in the \citet{Schlegel1998} reddening derivation. This must be added to the scatter in the Monte Carlo models because of the relatively small number of photons at high latitudes and to the small range in the fluxes to be fitted. Given these uncertainties, it appears that the data are consistent with an $a = 0.36$ and  $g = 0.5$ but a better understanding of the statistics is needed before a firm conclusion may be drawn.

The offset between the model predictions and the data is much better defined at a level of about 300 and 600 \photu\ in the FUV and the NUV, respectively (Fig. 19). The offsets are consistent between both poles at both bands. There have been a number of measurements of the diffuse Galactic light at high latitudes which have determined the offset at zero reddening to be about 300 \photu\ \citep{Anderson1979, Paresce1979, Paresce1980, Zvereva1982, Joubert1983, Jakobsen1984, Tennyson1988,  Onaka1989, Hamden2013}, which is generally thought to be due to an extragalactic background (see \citet{Bowyer1991, Henry1991} for discussion and references). \citep{Murthy2014a} derived airglow contributions of 220 \photu\ in the FUV and 350 \photu\ in the NUV using a different (empirical) method of analysis of the \galex\ data, which would leave a residual of about 100 and 250 \photu\ in the FUV and NUV, respectively. It is difficult to separate airglow from the other contributors without spectral diagnostics and I will defer a self-consistent determination of the offsets for a future work.

\begin{table*}
\centering
\caption{Optical constants and offsets in the literature.}
\label{ag_table}
\begin{tabular}{lllll}
\hline
\hline
References & Wavelength & \emph{a} & \emph{g} & Offset\\
		   & \AA & & & \photu\\
\hline

\cite{Witt1973}     & 1500         & >0.6             & <0.5       & -   \\
\cite{Lillie1976}   & 3000         & 0.7$\pm$0.1      & 0.6--0.9   & -   \\
 				    & 2200         & 0.35$\pm$0.05    & 0.6--0.9   & -   \\
 				    & 1550         & 0.6$\pm$0.05     & 0.6--0.9   & -   \\
\cite{Morgan1976}   & 2740         & 0.65$\pm$0.1     & 0.75       & -   \\
\cite{Morgan1978}   & 2740         & 0.68             & 0.5        & -   \\
 				    & 2350         & 0.51 			  & 0.5 	   & -   \\
 				    & 1950         & 0.53 			  & 0.5 	   & -   \\
 				    & 1550         & 0.5 			  & 0.5 	   & -   \\
\cite{Anderson1979} & 1230--1680   & - 	              & - 		   & $285\pm32$\\
\cite{Paresce1980}  & 1350--1550   & 0.5 	          & 0.5 	   & <300 \\
\cite{Feldman1981}  & 1200--1670   & - 	              &	-          & $150\pm50$\\
\cite{Henry1981}    & 1565         & >0.5 			  & >0.7 	   & -  \\
\cite{Witt1982}     & 1400         & 0.6 			  & 0.25 	   & -  \\
 				    & 2000         & 0.42 			  & 0.4 	   & -  \\
\cite{Joubert1983}  & 1690--2200   & - 		          & 0.6--0.7   & -  \\
\cite{Holberg1986}  & 500--1200    & - 		          & 		   & 100--200\\
\cite{Tennyson1988} & 1700--2850   & -                & -          & $300\pm100$\\
\cite{Fix1989} 		& 1500 		   & - 	              & >0.9	   & $530\pm80$\\
\cite{Hurwitz1991}  & 1415--1835   & 0.13--0.24       & <0.4       & 50  \\
\cite{Murthy1991}   & 912--1216    & <0.1 	          & >0.95      & - \\
\cite{Onaka1991}    & 1500         & $\geq$0.32       & $\geq$0.5  & 200--300  \\
\cite{Witt1992}     & 1400--2200   & 0.65             & 0.75       & -  \\
\cite{Henry1993}    & 1500         & 0.5              & 0.7        & 300$\pm$100  \\
\cite{Witt1993}     & 1000--1600   & 0.42$\pm$0.04    & 0.75       & -  \\
\cite{Murthy1993a}  & 912--1150    & >0.3             & <0.8       & -  \\
\cite{Murthy1993b}  & 1100--1860   & 0.5--0.7         & -          & - \\
\cite{Gordon1994}   & 1362         & 0.47--0.7        & <0.8       & -  \\
                    & 1769         & 0.55--0.72       & <0.8       & -\\
\cite{Hurwitz1994}  & 1600         & 0.6$\pm$0.1      & 0.5$\pm$0.15 & -  \\
\cite{Witt1994}     & 1500         & 0.5              & 0.9        & -  \\
\cite{Calzetti1995} & 1200--1600   & 0.7--0.8         & 0.75$\pm$0.05 & -  \\
	                & 2300         & 0.4              & 0.6        & -  \\
\cite{Murthy1995}   & 1250--2000   & 0.3--0.6         & -          & 100--400 \\
\cite{Sasseen1996}  & 1400--1800   & 0.3              & 0.8        & -  \\
\cite{Witt1997}     & 1400--1800   & 0.45$\pm$0.05    & 0.68$\pm$0.1 & $160\pm50$  \\
\cite{Schiminovich2001} & 1740     & 0.45$\pm$0.05    & 0.77$\pm$0.1 & 200$\pm$100 \\
\cite{Burgh2002}    & 900--1400    & 0.2--0.4         & 0.85 \\
\cite{Henry2002}    & 1500         & 0.1              & -          & -  \\
\cite{Mathis2002}   & 1300         & $\geq$0.5        & 0.6--0.85 & - \\
\cite{Gibson2003}   & <2600        & 0.22$\pm$0.07    & 0.74$\pm$0.06 & - \\
\cite{Weller1983}   & 1220--1500   & -                & -          & 200--300\\
\cite{Shalima2004}  & 1100         & 0.4$\pm$0.2      & -          & -  \\
\cite{Sujatha2005}  & 1100         & 0.4$\pm$0.1      & 0.55$\pm$0.25 & - \\
\cite{Shalima2006}  & 900--1200    & 0.3--0.7         & 0.55--0.85 & - \\
\cite{Sujatha2007}  & 900--1200    & 0.28$\pm$0.04    & 0.61$\pm$0.07 & -  \\	
\cite{Lee2008}      & 1370--1670   & 0.36$\pm$0.2     & 0.52$\pm$0.22 & \\
\cite{Sujatha2009}  & 1350--1750   & 0.4              & 0.7        & - \\
 				    & 1750--2850   &                  &            &   \\
\cite{Puthiyaveettil2010} & 1400--1900 & 0.6          & 0.8        & 500 \\
\cite{Sujatha2010}  & 1350--1750   & 0.32$\pm$0.09    & 0.51$\pm$0.19 & 30$\pm$10  \\
                    & 1750--2850   & 0.45$\pm$0.08    & 0.56$\pm$0.10 & 49$\pm$13\\
\cite{Murthy2011}   & 1521         & -                & 0.58$\pm$0.12 & -  \\
                    & 2320         &                  & 0.72$\pm$0.06& \\
\cite{Jo2012}       & 1350--1750   & 0.39$\pm$0.45    & 0.45$\pm$0.2 & -  \\
\cite{Choi2013}     & 1330--1780   & 0.38$\pm$0.06    & 0.46$\pm$0.06 & - \\
\cite{Hamden2013}   & 1344--1786   & 0.62$\pm$0.04    & 0.78$\pm$0.05 & 300  \\
\cite{Lim2013}      & 1360--1680   & 0.42$\pm$0.05    & 0.20--0.58  & -  \\
\cite{Jyothy2015}   & 1521         & 0.6--0.7         & 0.2--0.4 & -  \\
                    & 2317         &                  &          &  \\

\hline
\end{tabular}
\end{table*}

\section{Summary}

I have presented a Monte Carlo model for calculating the dust scattered starlight over the entire Galaxy. The main conclusions are as follows:
\begin{enumerate}

\item A multiple scattering model increases the scattered flux by about 30\%\ over the single scattering approximation regardless of the optical depth.

\item The total scattered flux is proportional to the square of the albedo and is greatest for isotropically scattering grains.

\item 90\% of the diffuse flux originates from less than 1000 stars and ~25\%\ from only 10 stars.

\item About half of the diffuse radiation seen at the Earth is scattered within 200 pc of the Sun with no radiation arising from further than 600 pc away. Most photons travel less than 200 pc before another interaction.

\item The all-sky diffuse radiation is fit well with $0.3 < a < 0.5$ and $g < 0.6$. The albedo is constrained by the total flux while $g$ is constrained by the amount of scattering far from bright stars.

\item The model predictions are close to the observed values at low and mid-latitudes for low optical depths. The fit is poorer at larger optical depths where the geometry is more complex.

\item There is reasonable agreement between the model ($a = 0.36; g = 0.5$) and the data at high latitudes but with considerable scatter.

\item There is an offset of 300 --- 700 \photu\ in both bands and at all latitudes which cannot be due to dust scattered radiation. \citet{Murthy2014a} estimated that the residual airglow was 220 \photu\ in the FUV and 350 \photu\ in the NUV implying that the offset at high latitudes is 100 and 250 \photu\ in the FUV and NUV bands, respectively. This is the component that is usually identified with extragalactic light \citep{Bowyer1991, Henry1991}. The offsets are larger at low latitudes and may be due to unaccounted sources such as molecular hydrogen emission or as yet undetermined sources \citep{Henry2015}.

\item I have tabulated results from the literature in Table \ref{ag_table}. In most cases, I have taken the results as specified by the authors which are difficult to translate into the  $1 \sigma$ results more often seen. In all cases, there is enough uncertainty in the data and the modelling that the formal limits are suggestive rather than definitive. There are a range of preferred values for the optical constants and the offset but I hope that the volume of data and improved modelling will yield tighter constraints on the dust properties.

\item I have released the software under a non-restrictive license \citep{Murthy2015} and have uploaded the model files to a public archive \citep{Murthy2016}. These files are the runs for Model 2 for a range of the optical constants and may be used for comparison with the diffuse background in the Milky Way. If an estimate of the diffuse flux is all that is needed, the file for $a = 0.4$ and $g = 0.6$ may be used at both 1500 \AA\ and 2300 \AA.

\end{enumerate}

\section{Further Work}

Although the overall fits are encouraging, there are a number of questions raised by the differences between the model and the observations. The local geometry of the exciting stars and the scattering dust are important in determining the diffuse background over much of the sky and their effects can be seen in Fig. \ref{obs_stars} where there are extended halos around bright stars such as Spica. There have been important new studies of the 3-dimensional distribution of the dust, most recently by \citet{Green2015}, which I will implement. It is likely that, at least in some parts of the sky, observations of the scattering will be better able to constrain the distance and density of the dust clouds than the standard extinction methods \citep{Lee2006}.

One of the major constraints in this work is the noise intrinsic to Monte Carlo modelling which can only be reduced by increasing the number of photons. Fortunately, Monte Carlo lend themselves well to modern HPC (high-performance computing) methods as well as processing on the GPU (graphics processing unit) and the next step is to port the software to that environment.

\section*{Acknowledgements}
I am grateful for a thorough review by the referee.
Part of this research has been supported by the Department of Science and Technology under Grant IR/S2/PU-006/2012. Some of the work was done while I was a Visiting Professor at Copperbelt University in Kitwe, Zambia. This research has made use of NASA's Astrophysics Data System Bibliographic Services. I have used the GnuDataLanguage (http://gnudatalanguage.sourceforge.net/index.php) for the analysis of this data. "The data presented in this paper were obtained from the Mikulski Archive for Space Telescopes (MAST). STScI is operated by the Association of Universities for Research in Astronomy, Inc., under NASA contract NAS5-26555. Support for MAST for non-HST data is provided by the NASA Office of Space Science via grant NNX09AF08G and by other grants and contracts."  





\bsp	
\label{lastpage}
\end{document}